\documentclass[twocolumn, journal]{IEEEtran}
\IEEEoverridecommandlockouts

\usepackage{color}
\usepackage{bm}
\usepackage{graphicx}
\usepackage{amsmath}
\usepackage{amssymb}
\usepackage{algorithm}
\usepackage{algorithmic}
\usepackage{multirow}
\usepackage{booktabs}
\usepackage{array}
\usepackage{amsthm}
\usepackage{lipsum}
\usepackage{enumerate}
\usepackage{stfloats}
\usepackage{subfigure}
\usepackage{cases}
\usepackage{diagbox}
\usepackage{threeparttable}
\usepackage{cite}
\usepackage{url}
\usepackage[OT1]{fontenc}

\newcommand{\non}{\nonumber}

\title{Non-Reciprocal Beyond Diagonal RIS:\\ Multiport Network Models and Performance Benefits in Full-Duplex Systems}

\author{Hongyu Li,~\IEEEmembership{Member,~IEEE} and Bruno Clerckx,~\IEEEmembership{Fellow,~IEEE}
\thanks{Manuscript received 6 November, 2024; revised 3 March, 2025; accepted 30 April, 2025. The associate editor coordinating the review of this article and approving it for publication was Prof. Liang Yang. This work has been partially supported by UKRI grant EP/Y004086/1, EP/X040569/1, EP/Y037197/1, EP/X04047X/1, EP/Y037243/1.}
\thanks{H. Li is with the Internet of Things Thrust, The Hong Kong University of Science and Technology (Guangzhou), Guangzhou 511400, China (e-mail: hongyuli@hkust-gz.edu.cn).}
\thanks{B. Clerckx is with the Department of Electrical and Electronic Engineering, Imperial College London, London SW7 2AZ, U.K., and also with Kyung Hee University, Seoul, Korea (e-mail: b.clerckx@imperial.ac.uk).}
\thanks{Corresponding author: Bruno Clerckx.}
}

\begin{document}

\maketitle
\thispagestyle{empty}
\begin{abstract}
    Beyond diagonal reconfigurable intelligent surface (BD-RIS) is a new advance in RIS techniques that introduces reconfigurable inter-element connections to generate scattering matrices not limited to being diagonal. BD-RIS has been recently proposed and proven to have benefits in enhancing channel gain and enlarging coverage in wireless communications. 
    Uniquely, BD-RIS enables reciprocal and non-reciprocal architectures characterized by symmetric and non-symmetric scattering matrices. However, the performance benefits and new use cases enabled by non-reciprocal BD-RIS for wireless systems remain unexplored. 
    This work takes a first step toward closing this knowledge gap and studies the non-reciprocal BD-RIS in full-duplex systems and its performance benefits over reciprocal counterparts. 
    We start by deriving a general RIS aided full-duplex system model using a multiport circuit theory, followed by a simplified channel model based on physically consistent assumptions. With the considered channel model, we investigate the effect of BD-RIS non-reciprocity and identify the theoretical conditions for reciprocal and non-reciprocal BD-RISs to simultaneously achieve the maximum received power of the signal of interest in the uplink and the downlink. Simulation results validate the theories and highlight the significant benefits offered by non-reciprocal BD-RIS in full-duplex systems. The significant gains are achieved because of the non-reciprocity principle which implies that if a wave hits the non-reciprocal BD-RIS from one direction, the surface behaves differently than if it hits from the opposite direction. This enables an uplink user and a downlink user at different locations to optimally communicate with the same full-duplex base station via a non-reciprocal BD-RIS, which would not be possible with reciprocal surfaces.
\end{abstract}


\section{Introduction}
\label{sec:intro}

As an energy-efficient passive planar surface which could be flexibly deployed in wireless propagation environments, reconfigurable intelligent surface (RIS) has been proven to have benefits in enhancing the communication quality and enlarging the wireless coverage in various scenarios \cite{di2020smart,wu2021intelligent,bjornson2020reconfigurable}. 
Based on microwave engineering theory \cite{pozar2021microwave}, RIS can be generally modeled as a large number of scattering elements connected to a multiport reconfigurable impedance network \cite{shen2021}. 
Most of the existing literature focuses on the use of diagonal (D) RIS, in which each port of the reconfigurable impedance network is connected to its own impedance to ground, mathematically leading to a diagonal scattering matrix \cite{shen2021}.
Benefiting from the flexible circuit topology design of the reconfigurable impedance network, a new advance in RIS techniques has been recently proposed, namely beyond diagonal (BD) RIS \cite{li2023reconfigurable}. 
The concept of BD-RIS is proposed by inter-connecting ports with additional reconfigurable impedance components, which mathematically generates scattering matrices not limited to being diagonal. 
This enables waves impinging on BD-RIS to travel through the surfaces, hence offering a new degree of freedom for wave manipulation in the analog RF domain. As such, BD-RIS includes D-RIS as a special case, while it goes beyond D-RIS to enable smarter beam manipulations to the wireless signals, and thus provide significant performance gains \cite{li2023reconfigurable,shen2021,nerini2024beyond}.

An $N_I$-element BD-RIS with an $N_I$-port lossless reconfigurable impedance network is generally characterized by a unitary scattering matrix $\mathbf{\Theta}$, $\mathbf{\Theta}^\mathsf{H}\mathbf{\Theta} = \mathbf{I}_{N_I}$. 
In addition, based on if the reconfigurable impedance network is reciprocal or not, BD-RIS can be categorized as two branches, namely, \textit{reciprocal} BD-RIS and \textit{non-reciprocal} BD-RIS, resulting in mathematically symmetric matrices with $\mathbf{\Theta} = \mathbf{\Theta}^\mathsf{T}$, or asymmetric scattering matrices with $\mathbf{\Theta} \ne \mathbf{\Theta}^\mathsf{T}$. 
Based on the above classification, D-RIS is a special case of reciprocal BD-RIS with $\mathbf{\Theta}$ being diagonal and symmetric, while non-reciprocity of the reconfigurable impedance network is an additional degree of freedom arising from BD-RIS\footnote{Conventional D-RIS are always reciprocal since the scattering matrix is a simple diagonal phase shift matrix that is symmetric, hence non-reciprocity can only be enabled by BD-RIS.}.
Reciprocal BD-RIS has been studied covering perspectives in modeling and architecture design, such as designing the circuit topologies to generate group/fully/forest/tree-connected architectures \cite{shen2021,nerini2024beyond}, in mode analysis \cite{li2023reconfigurable}, in beamforming design \cite{nerini2023closed,zhou2023optimizing}, and in hardware impairment analysis \cite{li2024beyond}.   
Moreover, the performance benefits of reciprocal BD-RIS has been shown in various scenarios, such as wideband communications \cite{demir2024wideband}, wireless power transfer \cite{azarbahram2025beyond}, non-terrestrial networks \cite{khan2024beyond}.
Non-reciprocal BD-RIS, however, is less studied. 
There has been one representative work focusing on the modeling and implementation of a specific model, whose scattering matrix is non-diagonal and asymmetric with $N_I$ nonzero off-diagonal entries \cite{li2022reconfigurable}. 
This work is further generalized to a non-diagonal group-connected design \cite{li2024coordinated}, which provides additional flexibility than the model in \cite{li2022reconfigurable} in wave manipulation due to the increasing number of nonzero entries of the scattering matrix.

To find the scenarios where non-reciprocal BD-RIS gives high performance and unique advantages over reciprocal BD-RIS and D-RIS, one interesting direction has recently been discovered in \cite{wang2024channel}. 
In this work, the non-reciprocal BD-RIS proposed in \cite{li2022reconfigurable} has been proven to have benefits over D-RIS in channel attack by breaking the reciprocity of RIS-aided uplink and downlink channels. 
More specifically, if a wave hits the non-reciprocal BD-RIS from one direction, the surface behaves differently than if it hits from the opposite direction. This property potentially allows the non-reciprocal BD-RIS to generate asymmetric beams for uplink and downlink communications.
Inspired by this work, another meaningful potential direction envisioned in \cite{youtube} is thus to analyze the behavior of reciprocal and non-reciprocal BD-RISs in wireless full-duplex systems \cite{zhang2016full,sabharwal2014band,ahmed2013rate}, which enable simultaneous uplink and downlink transmissions and are thus sensitive to the wireless channel reciprocity. Note that the use of conventional D-RIS in full-duplex systems has been investigated from various perspectives, such as beamforming design \cite{guan2022joint}, resource allocation \cite{xu2020resource}, and interference management \cite{zhang2023ris}. Nevertheless, the impact of wireless channel non-reciprocity in full-duplex systems has never been explored and is only visible when non-reciprocal BD-RIS is applied.

Based on the above considerations, in this paper, we identify a first scenario where gains arising from non-reciprocity in BD-RIS can be very significant. This further leads to the following contributions. 

\textit{First}, we use multiport network theory to derive the general RIS aided full-duplex system model, which consists of a multi-antenna full-duplex base station, an RIS, an uplink multi-antenna user, and a downlink multi-antenna user. We also derive its simplified version based on physically consistent assumptions, where the impact of self-interference at the full-duplex base station is highlighted. 

\textit{Second}, in the derived system model, we analyze the impact of the structural scattering at BD-RIS, which refers to the ``virtual'' direct link between the base station and the user constructed by BD-RIS when it is turned OFF, i.e., its scattering matrix equals zero.  
Departing from most existing RIS literature \cite{di2020smart,wu2021intelligent} where the structure scattering is simply ignored, we reach to a more accurate channel model and explicitly study its impact in full-duplex systems.

\textit{Third}, based on the proposed general model, we consider a simple BD-RIS aided full-duplex system, which consists of a single-antenna full-duplex base station, a BD-RIS, a single-antenna uplink user and a single-antenna downlink user. 
In this scenario, we derive the theoretical conditions for both reciprocal and non-reciprocal BD-RISs to simultaneously maximize the received powers of the signal of interest in the uplink and downlink. 

\textit{Fourth}, we provide simulation results to verify the theoretical derivations. More importantly, we visualize the best choice for locating the base station and the users, such that applying the non-reciprocal BD-RIS has the most significant performance benefits in full-duplex systems.
Our results demonstrate the uniqueness of non-reciprocal BD-RIS in full duplex systems. The asymmetric beams that the non-reciprocal BD-RIS generates for uplink and downlink enable an uplink user and a downlink user at different locations to optimally communicate with the same full-duplex base station. This would not be possible with a reciprocal surface unless the two users were aligned, which explains the observed significant performance gains offered by non-reciprocal BD-RIS.

\textit{Organization:} Section II introduces the multiport network analysis and classifies the RIS based on properties of the scattering matrix. Section III derives the RIS aided full-duplex system model. Section IV studies the theoretical conditions that non-reciprocal BD-RIS outperforms reciprocal BD-RIS in full-duplex systems. 
Section V numerically verifies the theoretical results and visualizes the benefit of non-reciprocal BD-RIS. Section VI concludes this work.

\textit{Notations:}
Boldface lower- and upper-case letters indicate column vectors and matrices, respectively. 
$(\cdot)^\mathsf{T}$, $(\cdot)^*$, $(\cdot)^\mathsf{H}$, and $(\cdot)^{-1}$ denote the transpose, conjugate, conjugate-transpose, and inverse operations, respectively.
$\mathbb{C}$ and $\mathbb{R}$ denote the sets of complex and real numbers, respectively. 
$\mathbb{E}$ denotes the statistical expectation. 
$\Re\{\cdot\}$ denotes the real part of complex numbers. 
$\mathsf{blkdiag}(\cdot)$ represents a block-diagonal matrix and $\mathsf{diag}(\cdot)$ represents a diagonal matrix.
$\|\cdot\|_2$ and $\|\cdot\|_\mathsf{F}$ denote the $\ell_2$ norm and Frobenius norm, respectively.
$\jmath=\sqrt{-1}$ denotes the imaginary unit.
$\mathsf{Tr}(\cdot)$ denotes the trace of a matrix.
$\mathbf{I}_M$ denotes an $M\times M$ identity matrix.
$\mathbf{0}_{M\times N}$ denotes an $M\times N$ all-zero matrix. 
$a\sim\mathcal{CN}(0,\sigma^2)$ characterizes the circular symmetric complex Gaussian distribution.
$[\mathbf{A}]_{i:i',j:j'}$ extracts the $i$-th to $i'$-th rows and the $j$-th to $j'$-th columns of $\mathbf{A}$. 
In microwave engineering, given an $N$-port network, its impedance parameter $\mathbf{Z}\in\mathbb{C}^{N\times N}$ links its scattering parameter $\mathbf{S}\in\mathbb{C}^{N\times N}$ by 
\begin{equation}
    \mathbf{S} = (\mathbf{Z} + Z_0\mathbf{I}_N)^{-1}(\mathbf{Z} - Z_0\mathbf{I}_N),
    \label{eq:s_z}
\end{equation}
where $Z_0$ denotes the reference impedance with $Z_0 = 50~\Omega$.

\section{Multiport Network Analysis}
\label{sec:syst_mod}

In this section, we take a systematic multiport network theoretical approach to model and analyze the RIS aided full-duplex system based on \cite{shen2021}.

\begin{figure*}
    \centering
    \includegraphics[width = 0.95\textwidth]{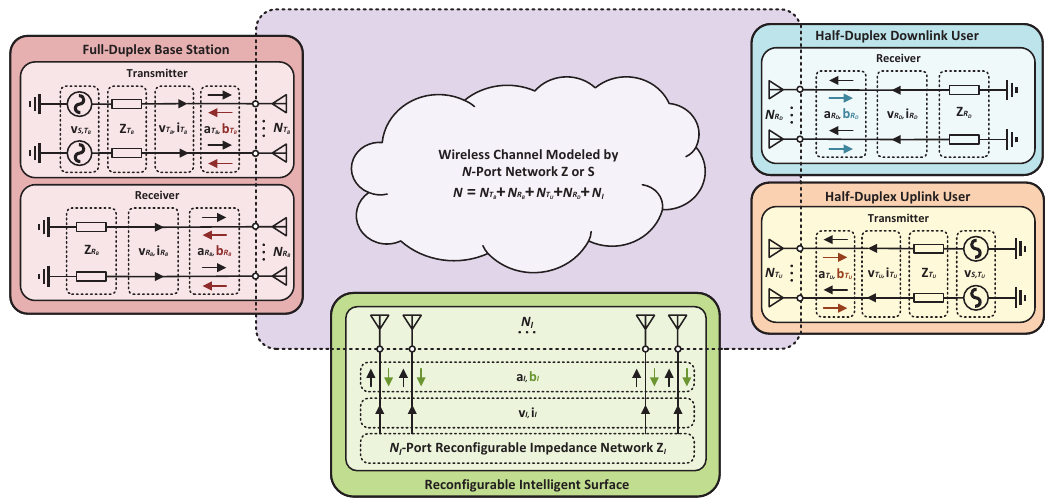}
    \caption{Diagram of RIS aided full-duplex systems.}
    \label{fig:network}
\end{figure*}

Consider an RIS aided multi-antenna full-duplex communication system consisting of a full-duplex base station, an RIS, an uplink half-duplex user, and a downlink half-duplex user, as illustrated in Fig. \ref{fig:network}\footnote{Our aim of considering a multi-antenna system here is to establish a general model including various scenarios, e.g., single-antenna systems, multi-antenna systems, and multiuser systems, which could provide modeling support for possible future works.}. Specifically, the full-duplex base station has $N_{T_B}$ transmitting antennas and $N_{R_B}$ receiving antennas; the RIS has $N_I$ elements; the uplink user has $N_{T_U}$ transmitting antennas; and the downlink user has $N_{R_D}$ receiving antennas.
The whole system is regarded as an $N = N_{T_B} + N_{T_U} + N_I + N_{R_B} + N_{R_D}$ port network characterized by its scattering matrix $\mathbf{S}\in\mathbb{C}^{N\times N}$ \cite{shen2021}, or, equivalently, its impedance matrix $\mathbf{Z}\in\mathbb{C}^{N\times N}$ \cite{gradoni2021end}, so that
\begin{equation}
    \mathbf{v} = \mathbf{Z}\mathbf{i}, ~~\mathbf{b} = \mathbf{S}\mathbf{a}, \label{eq:sz_nport}
\end{equation}
where the voltage vector $\mathbf{v} = [\mathbf{v}_{T_B};\mathbf{v}_{T_U};\mathbf{v}_I;\mathbf{v}_{R_B};\mathbf{v}_{R_D}]\in\mathbb{C}^{N\times 1}$, current vector $\mathbf{i} = [\mathbf{i}_{T_B};\mathbf{i}_{T_U};\mathbf{i}_I;\mathbf{i}_{R_B};\mathbf{i}_{R_D}]\in\mathbb{C}^{N\times 1}$, reflected wave vector $\mathbf{b} = [\mathbf{b}_{T_B};\mathbf{b}_{T_U};\mathbf{b}_I;\mathbf{b}_{R_B};\mathbf{b}_{R_D}]\in\mathbb{C}^{N\times 1}$, and incident wave vector $\mathbf{a} = [\mathbf{a}_{T_B};\mathbf{a}_{T_U};\mathbf{a}_I;\mathbf{a}_{R_B};\mathbf{a}_{R_D}]\in\mathbb{C}^{N\times 1}$ collect the voltages, currents, reflected waves, and incident waves at the ports of transmitters, RIS, and receivers, as illustrated in Fig. \ref{fig:network}.
The voltages, currents, incident and reflected waves can relate to each other by \cite{nerini2024universal}
\begin{equation}
    \mathbf{v} = \mathbf{a} + \mathbf{b}, ~~\mathbf{i} = \frac{\mathbf{a}-\mathbf{b}}{Z_0}. \label{eq:viab}
\end{equation}

\subsection{Transmitters and Receivers}

The transmitters at the base station and at the uplink user are respectively characterized by the source impedance matrices $\mathbf{Z}_{T_B}\in\mathbb{C}^{N_{T_B}\times N_{T_B}}$ and $\mathbf{Z}_{T_U}\in\mathbb{C}^{N_{T_U}\times N_{T_U}}$ or, equivalently, the scattering matrices $\mathbf{\Gamma}_{T_B}\in\mathbb{C}^{N_{T_B}\times N_{T_B}}$ and $\mathbf{\Gamma}_{T_U}\in\mathbb{C}^{N_{T_U}\times N_{T_U}}$, as shown in Fig. \ref{fig:network}. Therefore, we have 
\begin{subequations}\label{eq:tx}
    \begin{align}
    &\mathbf{v}_{T_B} = \mathbf{v}_{S,T_B} - \mathbf{Z}_{T_B}\mathbf{i}_{T_B}, ~ \mathbf{v}_{T_U} = \mathbf{v}_{S,T_U} - \mathbf{Z}_{T_U}\mathbf{i}_{T_U},\\
    &\mathbf{a}_{T_B} = \mathbf{b}_{S,T_B} + \mathbf{\Gamma}_{T_B}\mathbf{b}_{T_B}, ~ \mathbf{a}_{T_U} = \mathbf{b}_{S,T_U} + \mathbf{\Gamma}_{T_U}\mathbf{b}_{T_U},
    \end{align}
\end{subequations}
where $\mathbf{v}_{S,T_B}\in\mathbb{C}^{N_{T_B}\times 1}$,  $\mathbf{v}_{S,T_U}\in\mathbb{C}^{N_{T_U}\times 1}$, $\mathbf{b}_{S,T_B}\in\mathbb{C}^{N_{T_B}\times 1}$, and $\mathbf{b}_{S,T_U}\in\mathbb{C}^{N_{T_U}\times 1}$, respectively, refer to the source voltage vectors and source wave vectors at the corresponding transmitters. 

The receivers at the base station and at the downlink user are respectively characterized by the load impedance matrices $\mathbf{Z}_{R_B}\in\mathbb{C}^{N_{R_B}\times N_{R_B}}$ and $\mathbf{Z}_{R_D}\in\mathbb{C}^{N_{R_D}\times N_{R_D}}$, or equivalently, the scattering matrices $\mathbf{\Gamma}_{R_B}\in\mathbb{C}^{N_{R_B}\times N_{R_B}}$ and $\mathbf{\Gamma}_{R_D}\in\mathbb{C}^{N_{R_D}\times N_{R_D}}$, as shown in Fig. \ref{fig:network}. Therefore, we have 
\begin{subequations}\label{eq:rx}
    \begin{align}
    &\mathbf{v}_{R_B} = - \mathbf{Z}_{R_B}\mathbf{i}_{R_B}, ~ \mathbf{v}_{R_D} = - \mathbf{Z}_{R_D}\mathbf{i}_{R_D},\\
    &\mathbf{a}_{R_B} = \mathbf{\Gamma}_{R_B}\mathbf{b}_{R_B}, ~ \mathbf{a}_{R_D} = \mathbf{\Gamma}_{R_D}\mathbf{b}_{R_D}.
    \end{align}
\end{subequations}

\subsection{Reconfigurable Intelligent Surface}

The RIS is characterized by the impedance matrix $\mathbf{Z}_I\in\mathbb{C}^{N_I\times N_I}$ of the reconfigurable impedance network, or equivalently, the scattering matrix $\mathbf{\Theta}\in\mathbb{C}^{N_I\times N_I}$. Therefore, we have 
\begin{equation}
    \mathbf{v}_I = -\mathbf{Z}_I\mathbf{i}_I, ~~\mathbf{a}_I = \mathbf{\Theta}\mathbf{b}_I.\label{eq:ris}
\end{equation}
Assuming lossless reconfigurable impedance network, we have a purely imaginary impedance matrix, i.e., $\Re\{\mathbf{Z}_I\} = \mathbf{0}_{N_I\times N_I}$, yielding $\mathbf{\Theta}^\mathsf{H}\mathbf{\Theta} = \mathbf{I}_{N_I}$ based on (\ref{eq:s_z})\footnote{According to microwave engineering \cite{pozar2021microwave}, a passive network means its net power is non-negative, i.e., the reflected power of the network is no larger than the incident power. The lossless network refers to an extreme case where the reflected power of the network is equal to the incident power. This implies the scattering matrix of the network is unitary.}. 
Further, we introduce the following two classification principles. 

\textit{Principle 1: Property of the Scattering Matrix.}
Based on whether the scattering matrix $\mathbf{\Theta}$ is diagonal or not, we can categorize RIS as follows.
\begin{itemize}
    \item D-RIS: Each port of the reconfigurable impedance network is connected to its own impedance component to ground without interacting with other ports, which yields a diagonal $\mathbf{Z}_I$, such that $\mathbf{\Theta}$ is diagonal by (\ref{eq:s_z}). The D-RIS has been well studied in existing literature \cite{di2020smart,wu2021intelligent}.
    \item BD-RIS: Part of/all the ports of the reconfigurable impedance networks are connected to each other, which generates $\mathbf{Z}_I$ and $\mathbf{\Theta}$ not limited to being diagonal. Architecture designs based on circuit topologies of the reconfigurable impedance network, mode analysis based on flexible antenna arrangements, and the performance benefits of BD-RIS over D-RIS based on efficient beamforming designs have been recently studied in \cite{shen2021,li2023reconfigurable,nerini2024beyond}.
\end{itemize}

\begin{figure}
    \centering
    \includegraphics[width=0.48\textwidth]{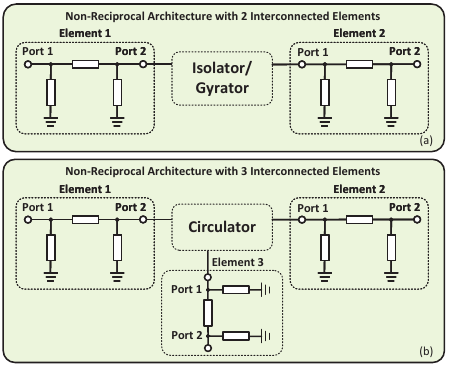}
    \caption{Illustrative examples of non-reciprocal architectures: (a) 2 elements interconnect
    with each other via an isolator or gyrator; (b) 3 elements interconnect with each other via a circulator.}
    \label{fig:nr_architecture}
\end{figure}

\textit{Principle 2: Reciprocity of the Reconfigurable Impedance Network.} 
Based on whether the reconfigurable impedance network is reciprocal or not, we can further categorize BD-RIS as follows.
\begin{itemize}
    \item Reciprocal BD-RIS: The reconfigurable impedance network is reciprocal, which yeilds $\mathbf{Z}_I = \mathbf{Z}_I^\mathsf{T}$ and thus $\mathbf{\Theta} = \mathbf{\Theta}^\mathsf{T}$ by (\ref{eq:s_z}). This could be potentially implemented by varactors and PIN diodes based on different circuit topologies \cite{shen2021}. 
    An extreme case where $\mathbf{\Theta}$ is diagonal corresponds to the D-RIS \cite{di2020smart,wu2021intelligent}.  
    \item Non-Reciprocal BD-RIS: The reconfigurable impedance network is non-reciprocal, which yields $\mathbf{Z}_I \ne \mathbf{Z}_I^\mathsf{T}$ and thus $\mathbf{\Theta} \ne \mathbf{\Theta}^\mathsf{T}$ by (\ref{eq:s_z}). This could be potentially implemented by non-reciprocal circuits, e.g., isolators, gyrators, circulators \cite{xu2024non,wang2024beyond}. Specifically, \cite{xu2024non,wang2024beyond} have provided two illustrative examples for a 2-element interconnected architecture realized by the isolator/gyrator and a 3-element interconnected architecture realized by the circulator, which are respectively given in Fig. \ref{fig:nr_architecture}(a) and Fig. \ref{fig:nr_architecture}(b).
    In addition, an extreme case is that $\mathbf{\Theta}$ has only $N_I$ nonzero entries occupying off-diagonal entries of the scattering matrix \cite{li2022reconfigurable}.
\end{itemize}

\textit{Remark 1:} We should highlight here that in BD-RIS, the symmetric constraints of $\mathbf{\Theta}$ can be accounted for or relaxed, hence leading to reciprocal and non-reciprocal BD-RIS, while D-RIS is a special case of the reciprocal BD-RIS. This implies that non-reciprocity is an additional degree of freedom arising from BD-RIS. 
The effect of BD-RIS reciprocity in full-duplex communication systems will be soon analyzed based on the following channel models. 

\section{RIS Aided Full-Duplex Systems}

In this section, we derive the general channel model for the RIS aided full-duplex system based on the multiport network analysis and simplify it based on physically consistent assumptions.

\subsection{General Channel for RIS Aided Full-Duplex Systems}

Defining $N_T = N_{T_B} + N_{T_U}$ and $N_R = N_{R_B} + N_{R_D}$, we have the wireless channel for RIS aided full-duplex systems given by $\mathbf{H}\in\mathbb{C}^{N_R\times N_T}$. Specifically, $\mathbf{H}$ relates the voltage at the transmitters, $\mathbf{v}_T = [\mathbf{v}_{T_B};\mathbf{v}_{T_U}]\in\mathbb{C}^{N_T\times 1}$, and the voltage at the receivers, $\mathbf{v}_R = [\mathbf{v}_{R_B};\mathbf{v}_{R_D}]\in\mathbb{C}^{N_R\times 1}$, through
\begin{equation}
    \mathbf{v}_R = \mathbf{H}\mathbf{v}_T.
\end{equation}
To obtain $\mathbf{H}$ as functions of $\mathbf{S}$, $\mathbf{\Gamma}_R = \mathsf{blkdiag}(\mathbf{\Gamma}_{R_B}, \mathbf{\Gamma}_{R_D})$, $\mathbf{\Gamma}_T = \mathsf{blkdiag}(\mathbf{\Gamma}_{T_B},\mathbf{\Gamma}_{T_U})$, and $\mathbf{\Theta}$, we block $\mathbf{Z}$ and $\mathbf{S}$ as 
\begin{equation}
    \mathbf{Z} = \left[\begin{matrix}
        \mathbf{Z}_{TT} &\mathbf{Z}_{TI} &\mathbf{Z}_{TR}\\
        \mathbf{Z}_{IT} &\mathbf{Z}_{II} &\mathbf{Z}_{IR}\\
        \mathbf{Z}_{RT} &\mathbf{Z}_{RI} &\mathbf{Z}_{RR}
    \end{matrix}\right],
    \mathbf{S} = \left[\begin{matrix}
        \mathbf{S}_{TT} &\mathbf{S}_{TI} &\mathbf{S}_{TR}\\
        \mathbf{S}_{IT} &\mathbf{S}_{II} &\mathbf{S}_{IR}\\
        \mathbf{S}_{RT} &\mathbf{S}_{RI} &\mathbf{S}_{RR}
    \end{matrix}\right],\label{eq:block}
\end{equation} 
and solve equations (\ref{eq:sz_nport})-(\ref{eq:ris}) following the steps in \cite{shen2021,nerini2024universal}. As such, the general RIS aided wireless channel $\mathbf{H}$ is given by 
\begin{equation}
    \mathbf{H} = (\mathbf{\Gamma}_R + \mathbf{I}_{N_R})\mathbf{T}_{RT}(\mathbf{I}_{N_T} + \mathbf{\Gamma}_T\mathbf{T}_{TT} + \mathbf{T}_{TT})^{-1},\label{eq:general_h}
\end{equation}
where 
\begin{subequations}
    \begin{align} 
        &\mathbf{T}_{RT} = [\mathbf{S}(\mathbf{I}_N - \mathbf{\Gamma}\mathbf{S})^{-1}]_{N_T + N_I + 1 : N,1 : N_T},\\ 
        &\mathbf{T}_{TT} = [\mathbf{S}(\mathbf{I}_N - \mathbf{\Gamma}\mathbf{S})^{-1}]_{1 : N_T,1 : N_T},
    \end{align}
\end{subequations}
with $\mathbf{\Gamma} = \mathsf{blkdiag}(\mathbf{\Gamma}_T,\mathbf{\Theta},\mathbf{\Gamma}_R)$.

Before going deeply into further analysis of the effect of RIS reciprocity in full-duplex systems, in the following subsection, we first make physically consistent assumptions to simplify the channel expression (\ref{eq:general_h}). 

\subsection{Simplified Channel for RIS Aided Full-Duplex Systems}

\textit{Assumption 1: Perfect Matching at Transmitters and Receivers.} The source impedance at the transmitters and the load impedance at the receivers are equal to the reference impedance $Z_0$, i.e., $\mathbf{Z}_{T_B} = Z_0\mathbf{I}_{N_{T_B}}$, $\mathbf{Z}_{T_U} = Z_0\mathbf{I}_{N_{T_U}}$, $\mathbf{Z}_{R_B} = Z_0\mathbf{I}_{N_{R_B}}$, and $\mathbf{Z}_{R_D} = Z_0\mathbf{I}_{N_{R_D}}$, yielding $\mathbf{\Gamma}_{T_B} = \mathbf{0}_{N_{T_B}\times N_{T_B}}$ and $\mathbf{\Gamma}_{T_U} = \mathbf{0}_{N_{T_U}\times N_{T_U}}$, $\mathbf{\Gamma}_{R_B} = \mathbf{0}_{N_{R_B}\times N_{R_B}}$, $\mathbf{\Gamma}_{R_D} = \mathbf{0}_{N_{R_D}\times N_{R_D}}$ based on (\ref{eq:s_z}). 
$\hfill\square$

\textit{Assumption 2: Unilateral Approximation \cite{ivrlavc2010toward}. } The unilateral approximation says that when there are large distances between the full-duplex base station, users, and the RIS, yielding large signal attenuation between devices, the electrical properties at the transmitter are approximately independent of those at the receiver. The unilateral approximation holds true when the minimum distance between devices, which is a function of the number of antennas, antenna spacing, and radiation pattern, is guaranteed \cite{ivrlavc2010toward}. 
As such, we assume $\mathbf{Z}_{IR} = \mathbf{0}_{N_I\times N_R}$ from the receivers to the RIS, $\mathbf{Z}_{TI} = \mathbf{0}_{N_T\times N_I}$ from the RIS to the transmitters,  $\mathbf{Z}_{T_BR_D} = [\mathbf{Z}_{TR}]_{1:N_{T_B},N_{R_B}+1:N_R} = \mathbf{0}_{N_{T_B}\times N_{R_D}}$ from the downlink user to the transmitter at the full-duplex base station, $\mathbf{Z}_{T_UR_B} = [\mathbf{Z}_{TR}]_{N_{T_B}+1:N_T,1:N_{R_B}} = \mathbf{0}_{N_{T_U}\times N_{R_B}}$ from the receiver at the full-duplex base station to the uplink user, and $\mathbf{Z}_{T_UR_D} = [\mathbf{Z}_{TR}]_{N_{T_B}+1:N_T,N_{R_B}+1:N_R} = \mathbf{0}_{N_{T_U}\times N_{R_D}}$ from the downlink user to the uplink user.
$\hfill\square$

\textit{Assumption 3: Perfect Matching with No Mutual Coupling at All Transmit and Receive Antennas, and All RIS Elements.} This mathematically means $\mathbf{Z}_{TT} = Z_0\mathbf{I}_{N_T}$, $\mathbf{Z}_{II} = Z_0\mathbf{I}_{N_I}$, $\mathbf{Z}_{RR} = Z_0\mathbf{I}_{N_R}$. $\hfill\square$

Assumptions 2 and 3 lead to, mathematically, the following impedance matrix of the $N$-port network:
\begin{equation}
    \mathbf{Z} = \left[\begin{matrix}
        Z_0\mathbf{I}_{N_T} &\mathbf{0} &\mathbf{Z}_{TR}\\
        \mathbf{Z}_{IT} &Z_0\mathbf{I}_{N_I} &\mathbf{0}\\
        \mathbf{Z}_{RT} &\mathbf{Z}_{RI} &Z_0\mathbf{I}_{N_R}
    \end{matrix}\right],
\end{equation} 
where $\mathbf{Z}_{TR} = \mathsf{blkdiag}(\mathbf{Z}_{T_BR_B},\mathbf{0}_{N_{T_U}\times N_{R_D}})$. 
In conventional half-duplex systems, $\mathbf{Z}_{TR}$ denotes the impedance from the receiver to the transmitter, and thus can be assumed to be zero by the unilateral approximation \cite{ivrlavc2010toward}. 
However, in the considered full-duplex systems, the nonzero term of $\mathbf{Z}_{TR}$, i.e., $\mathbf{Z}_{T_BR_B}$, denotes the impedance from the receiver to the transmitter within the same full-duplex device, which thus characterizes the so-called self-interference in conventional full-duplex systems \cite{zhang2016full,sabharwal2014band} due to the reciprocity of antennas and physical wireless channels. 
Together with Assumption 1 and (\ref{eq:s_z}), we have the following result.

\textit{Result 1:} By (\ref{eq:s_z}) and Assumptions 1-3, (\ref{eq:general_h}) is simplified as 
\begin{equation}
    \begin{aligned}
        \mathbf{H} &= (\mathbf{S}_{RT} + \mathbf{S}_{RI}(\mathbf{I}_{N_I} - \mathbf{\Theta}\mathbf{S}_{II})^{-1}\mathbf{\Theta}\mathbf{S}_{IT})\\
        &~~~\times(\mathbf{I}_{N_T} + \mathbf{S}_{TT} + \mathbf{S}_{TI}(\mathbf{I} - \mathbf{\Theta}\mathbf{S}_{II})^{-1}\mathbf{\Theta}\mathbf{S}_{IT})^{-1},
    \end{aligned}
    \label{eq:simplified_h}
\end{equation}
where the quantities in (\ref{eq:block}) can be written as
\begin{subequations}
    \begin{align}
        &\mathbf{S}_{TT} = -\frac{1}{2Z_0}\mathbf{Z}_{TR}\mathbf{A}\mathbf{Z}_{RT},
        \mathbf{S}_{TI} = -\frac{1}{2Z_0}\mathbf{Z}_{TR}\mathbf{A}\mathbf{Z}_{RI},\\
        &\mathbf{S}_{IT} = \mathbf{A}_{II}\mathbf{Z}_{IT} + \mathbf{A}_{IR}\mathbf{Z}_{RT},
        \mathbf{S}_{II} = \mathbf{A}_{IR}\mathbf{Z}_{RI}, \\
        &\mathbf{S}_{RT} = \mathbf{A}_{RI}\mathbf{Z}_{IT} + \mathbf{A}_{RR}\mathbf{Z}_{RT},
        \mathbf{S}_{RI} = \mathbf{A}_{RR}\mathbf{Z}_{RI},
    \end{align}
\end{subequations}
with  
\begin{subequations}
    \begin{align}
    \non
        \mathbf{A} &= \left[\begin{matrix}
            \mathbf{A}_{II} &\mathbf{A}_{IR}\\
            \mathbf{A}_{RI} &\mathbf{A}_{RR}
        \end{matrix}\right] \\
        &= \left[\begin{smallmatrix}
            \frac{1}{2Z_0}\mathbf{I}_{N_I} - \frac{1}{8Z_0^3}\mathbf{Z}_{IT}\mathbf{Z}_{TR}\mathbf{B}^{-1} \mathbf{Z}_{RI} &\frac{1}{4Z_0^2}\mathbf{Z}_{IT}\mathbf{Z}_{TR}\mathbf{B}^{-1}\\
            -\frac{1}{2Z_0}\mathbf{B}^{-1}\mathbf{Z}_{RI} &\mathbf{B}^{-1}
        \end{smallmatrix}\right], \\ 
        \mathbf{B} &= 2Z_0\mathbf{I}_{N_R} -\frac{1}{2Z_0}\mathbf{Z}_{RT}\mathbf{Z}_{TR} + \frac{1}{4Z_0^2}\mathbf{Z}_{RI}\mathbf{Z}_{IT}\mathbf{Z}_{TR}. 
    \end{align}
\end{subequations}
$\hfill\square$

The simplified channel model (\ref{eq:simplified_h}) is still complex with multiple inverses. 
Specifically, from Result 1, we observe that the presence of the matrix inverse $\mathbf{B}^{-1}$ makes it difficult to clearly map the scattering parameter blocks in $\mathbf{S}$ to the impedance parameter ones in $\mathbf{Z}$. 
This essentially comes from the nonzero term $\mathbf{Z}_{TR}$, which is often ignored in conventional full-duplex systems \cite{zhang2016full,sabharwal2014band}.
In other words, compared to the channel model used in conventional full-duplex systems \cite{zhang2016full,sabharwal2014band}, (\ref{eq:simplified_h}) is a more accurate version explicitly capturing the impact of self-interference at full-duplex base station. This thus motivates us to make the following assumption.

\textit{Assumption 4: Effective Self-Interference Suppression at the Propagation Domain \cite{zhang2016full,sabharwal2014band,everett2014passive,aryafar2012midu,everett2011empowering}}. 
The self-interference suppression techniques can be generally classified into three categories: wireless propagation domain, analog circuit domain, and digital domain techniques \cite{sabharwal2014band}. Specifically, the wireless propagation domain techniques aim to isolate the transmit chain from the receive chain of the full-duplex device, such that the electrical properties at the transmitter can be seen as independent of those at the receiver. When there are separate transmitting and reflecting antennas within the full-duplex device, this can be done by spacing antennas apart and placing absorptive shielding in between \cite{everett2014passive}, cross-polarization between transmitting and reflecting antennas \cite{aryafar2012midu}, and increasing antenna directionality \cite{everett2011empowering}. 
When the transmitter and the receiver in the full-duplex device share antennas, this can be done using the circulator.
As such, we assume $\mathbf{Z}_{T_BR_B} = \mathbf{0}_{N_{T_B}\times N_{R_B}}$ from the receiver to the transmitter in the full-duplex base station. $\hfill\square$

\textit{Result 2}: By Assumption 4, (\ref{eq:simplified_h}) reduces to 
\begin{equation}
    \label{eq:simplified_h1}
    \mathbf{H} = \mathbf{S}_{RT} + \mathbf{S}_{RI}\mathbf{\Theta}\mathbf{S}_{IT},
\end{equation}
where the quantities in (\ref{eq:block}) are given by
\begin{subequations}\label{eq:mappings}
    \begin{align}
        &\mathbf{S}_{TT} = \mathbf{0}_{N_T\times N_T},~~ \mathbf{S}_{TI} = \mathbf{0}_{N_T\times N_I}, \\ &\mathbf{S}_{IT} = \frac{1}{2Z_0}\mathbf{Z}_{IT}, ~~\mathbf{S}_{II} = \mathbf{0}_{N_I\times N_I},\\
        &\mathbf{S}_{RT} = \frac{1}{2Z_0}\mathbf{Z}_{RT} - \frac{1}{4Z_0^2}\mathbf{Z}_{RI}\mathbf{Z}_{IT}, ~\mathbf{S}_{RI} = \frac{1}{2Z_0}\mathbf{Z}_{RI}.
    \end{align}
\end{subequations}
$\hfill\square$

\begin{figure}
    \centering
    \includegraphics[width = 0.48\textwidth]{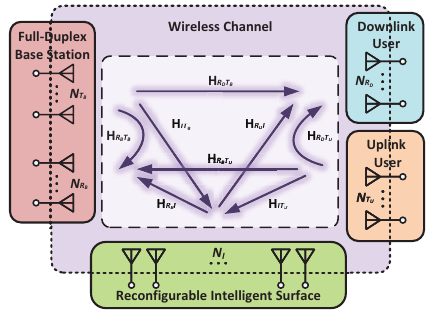}
    \caption{Illustration of channel blocks in (\ref{eq:simplified_h2}).}
    \label{fig:channel}
\end{figure}

Since the impedance parameters 
$\mathbf{Z}_{RT}$, $\mathbf{Z}_{RI}$, and $\mathbf{Z}_{IT}$ capture the open-circuit radiation patterns between transmitting and receiving antennas \cite{pozar2021microwave}, they essentially characterize the wireless channels between devices \cite{shen2021}.
To facilitate understanding, we introduce fresh notations 
\begin{equation}
    \mathbf{H}_{RT} = \frac{\mathbf{Z}_{RT}}{2Z_0}, ~ \mathbf{H}_{RI} = \frac{\mathbf{Z}_{RI}}{2Z_0}, ~ \mathbf{H}_{IT} = \frac{\mathbf{Z}_{IT}}{2Z_0},
\end{equation}
to denote channels between devices, and rewrite (\ref{eq:simplified_h1}) as
\begin{equation}
    \mathbf{H} = \mathbf{H}_{RT} + \mathbf{H}_{RI}(\mathbf{\Theta} - \mathbf{I}_{N_I})\mathbf{H}_{IT}.
\end{equation}
We further block $\mathbf{H}_{RT}$, $\mathbf{H}_{RI}$, and $\mathbf{H}_{IT}$ as 
\begin{equation}
    \begin{aligned}
    \mathbf{H}_{RT} &= \left[\begin{matrix}
            \mathbf{H}_{R_BT_B} &\mathbf{H}_{R_BT_U}\\
            \mathbf{H}_{R_DT_B} &\mathbf{H}_{R_DT_U}
        \end{matrix}\right], \\
        \mathbf{H}_{RI} &= \left[\begin{matrix}
        \mathbf{H}_{R_BI}\\ \mathbf{H}_{R_DI}
        \end{matrix}\right], ~~\mathbf{H}_{IT} = [\mathbf{H}_{IT_B} ~\mathbf{H}_{IT_U}],
    \end{aligned}
\end{equation}
where $\mathbf{H}_{R_BT_B}\in\mathbb{C}^{N_{R_B}\times N_{T_B}}$ denotes the self-interference channel at the full-duplex base station, $\mathbf{H}_{R_BT_U}\in\mathbb{C}^{N_{R_B}\times N_{T_U}}$ denotes the channel from the uplink user to base station, $\mathbf{H}_{R_DT_B}\in\mathbb{C}^{N_{R_D}\times N_{T_B}}$ denotes the channel from the base station to downlink user, $\mathbf{H}_{R_DT_U}\in\mathbb{C}^{N_{R_D}\times N_{T_U}}$ denotes the channel from the uplink user to downlink user, $\mathbf{H}_{R_BI}\in\mathbb{C}^{N_{R_B}\times N_{I}}$ denotes the channel from RIS to the base station, $\mathbf{H}_{R_DI}\in\mathbb{C}^{N_{R_D}\times N_{I}}$ denotes the channel from RIS to the downlink user, $\mathbf{H}_{IT_B}\in\mathbb{C}^{N_{I}\times N_{T_B}}$ denotes the channel from the base station to RIS, and $\mathbf{H}_{IT_U}\in\mathbb{C}^{N_{I}\times N_{T_U}}$ denotes the channel from the uplink user to RIS. 
To facilitate understanding, these channel blocks are illustrated in Fig. \ref{fig:channel}.
As such, $\mathbf{H}$ can be rewritten as
\begin{equation}\label{eq:simplified_h2}
        \mathbf{H} = \left[\begin{matrix}
            \bar{\mathbf{H}}_{R_BT_B} &\bar{\mathbf{H}}_{R_BT_U}\\
            \bar{\mathbf{H}}_{R_DT_B} &\bar{\mathbf{H}}_{R_DT_U}
        \end{matrix}\right],
\end{equation}
with diagonal blocks respectively referring to the interference channels at full-duplex base station and the channels from the uplink user to the downlink user, and off-diagonal blocks referring to the channels between the base station and the user, each of which are expressed as
\begin{subequations}\label{eq:channel_blocks}
    \begin{align}
        \bar{\mathbf{H}}_{R_BT_B} &= \underbrace{\mathbf{H}_{R_BT_B} + \mathbf{H}_{R_BI}(\mathbf{\Theta}-\mathbf{I}_{N_I})\mathbf{H}_{IT_B}}_{\text{self- and loop-interference channel at base station}},\\ \bar{\mathbf{H}}_{R_BT_U} &= \underbrace{\mathbf{H}_{R_BT_U} + \mathbf{H}_{R_BI}(\mathbf{\Theta}-\mathbf{I}_{N_I})\mathbf{H}_{IT_U}}_{\text{overall uplink user-base station channel}},\\
        \bar{\mathbf{H}}_{R_DT_B} &= \underbrace{\mathbf{H}_{R_DT_B} + \mathbf{H}_{R_DI}(\mathbf{\Theta}-\mathbf{I}_{N_I})\mathbf{H}_{IT_B}}_{\text{overall base station-downlink user channel}},\\ \bar{\mathbf{H}}_{R_DT_U} &= \underbrace{\mathbf{H}_{R_DT_U} + \mathbf{H}_{R_DI}(\mathbf{\Theta}-\mathbf{I}_{N_I})\mathbf{H}_{IT_U}}_{\text{overall uplink user-downlink user channel}}.
    \end{align}
\end{subequations}

\textit{Remark 2:} Note that the channel model (\ref{eq:simplified_h2}) aligns with that in \cite{nossek2024physically}, while it differs from that in most existing RIS literature \cite{wu2021intelligent,di2020smart}, i.e., 
\begin{equation}\label{eq:simplified_h2_noss}
    \begin{aligned}
        \tilde{\mathbf{H}} &= \mathbf{H}_{RT} + \mathbf{H}_{RI}\mathbf{\Theta}\mathbf{H}_{IT}\\
        &= \left[\begin{matrix}
            \underbrace{\mathbf{H}_{R_BT_B} +\mathbf{H}_{R_BI}\mathbf{\Theta}\mathbf{H}_{IT_B}}_{\text{self- and loop-interference channel}} & \underbrace{\mathbf{H}_{R_BT_U} + \mathbf{H}_{R_BI}\mathbf{\Theta}\mathbf{H}_{IT_U}}_{\text{uplink user-base station channel}}\\
            \underbrace{\mathbf{H}_{R_DT_B} +\mathbf{H}_{R_DI}\mathbf{\Theta}\mathbf{H}_{IT_B}}_{\text{base station-downlink user channel}} & \underbrace{\mathbf{H}_{R_DT_U}+\mathbf{H}_{R_DI}\mathbf{\Theta}\mathbf{H}_{IT_U}}_{\text{uplink user-downlink user channel}}
        \end{matrix}\right].
    \end{aligned}
\end{equation}
The above channel model (\ref{eq:simplified_h2_noss}) is obtained by simply setting $\mathbf{H}_{RT} = \mathbf{S}_{RT}$, $\mathbf{H}_{RI}=\mathbf{S}_{RI}$, and $\mathbf{H}_{IT} = \mathbf{S}_{IT}$, while the direct mapping from the wireless channel to the scattering parameter is not always physically compliant. More specifically, comparing (\ref{eq:simplified_h2}) and (\ref{eq:channel_blocks}) with (\ref{eq:simplified_h2_noss}), the physical meaning of the additional term, i.e., $-\mathbf{H}_{RI}\mathbf{H}_{IT}$, corresponds to the structural scattering, which refers to the ``virtual'' direct link constructed by RIS when all ports at the RIS are perfectly matched, i.e., $\mathbf{\Theta} = \mathbf{0}_{N_I\times N_I}$ \cite{nossek2024physically}.
The structural scattering intrinsically exists since even if all ports at the RIS are perfectly matched with $\mathbf{\Theta} = \mathbf{0}_{N_I\times N_I}$, antennas are still radiating because of the presence of non-zero current in the RIS elements/antennas, hence generating this additional term.
This structural scattering term is often ignored but will impact significantly the behavior of the BD-RIS in full-duplex systems, which will be detailed below.

\section{Case Study}
In this section, we adopt the simplified channel models (\ref{eq:simplified_h2}) and (\ref{eq:simplified_h2_noss}) in a BD-RIS aided full-duplex system with a full-duplex base station and two half-duplex users and analyze how the BD-RIS reciprocity affects the system performance. 

\subsection{System Model}

\begin{figure}
    \centering
    \includegraphics[width = 0.45\textwidth]{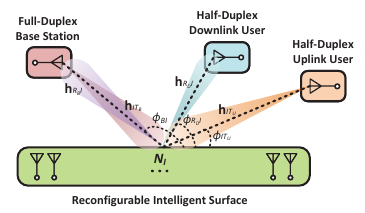}
    \caption{Transmission model with one single-antenna full-duplex base station, one RIS, and two full-duplex users.}
    \label{fig:case}
\end{figure}

We consider a BD-RIS aided full-duplex system with a single-antenna full-duplex base station (by single-antenna we assume $N_{T_B} = N_{R_B} = 1$ with the shared antenna), an $N_{I}$-element BD-RIS, and two single-antenna half-duplex users (one uplink user with $N_{T_U} = 1$ and one downlink user with $N_{R_D}=1$), as illustrated in Fig. \ref{fig:case}. 
In this case, we focus purely on the effect of RIS in full-duplex systems and make the following two assumptions. 

\textit{Assumption 5: Perfect Cancellation of the Self-Interference \cite{ahmed2013rate}.} In addition to the self-interference suppression in the wireless domain, cancellation techniques in the analog circuit and digital domain can be further adopted to effectively deal with the self-interference \cite{sabharwal2014band}. This allows us to assume $z_{R_BT_B} = [\mathbf{Z}_{RT}]_{1,1} = 0$. $\hfill\square$

\textit{Assumption 6: The Absence of Direct Links.} That is, $z_{R_BT_U} = [\mathbf{Z}_{RT}]_{1,2} = 0$,  $z_{R_DT_B} = [\mathbf{Z}_{RT}]_{2,1} = 0$, and $z_{R_DT_U} = [\mathbf{Z}_{RT}]_{2,2} = 0$. $\hfill\square$

\textit{Result 3:} By Assumptions 5 and 6, the structural scattering aware channel in (\ref{eq:simplified_h2}) writes as 
\begin{equation}\label{eq:simplified_h3}
    \begin{aligned}
        \mathbf{H} &= \mathbf{H}_{RI}(\mathbf{\Theta}-\mathbf{I}_{N_I})\mathbf{H}_{IT}\\
        &= \left[\begin{matrix}
            \underbrace{\mathbf{h}_{R_BI}^\mathsf{T}(\mathbf{\Theta}-\mathbf{I}_{N_I})\mathbf{h}_{IT_B}}_{\text{loop-interference channel}} & \underbrace{\mathbf{h}_{R_BI}^\mathsf{T}(\mathbf{\Theta}-\mathbf{I}_{N_I})\mathbf{h}_{IT_U}}_{\text{user-RIS-base station channel}}\\
            \underbrace{\mathbf{h}_{R_DI}^\mathsf{T}(\mathbf{\Theta}-\mathbf{I}_{N_I})\mathbf{h}_{IT_B}}_{\text{base station-RIS-user channel}} & \underbrace{\mathbf{h}_{R_DI}^\mathsf{T}(\mathbf{\Theta}-\mathbf{I}_{N_I})\mathbf{h}_{IT_U}}_{\text{user-RIS-user channel}}
        \end{matrix}\right],
    \end{aligned}
\end{equation}
where $\mathbf{h}_{R_DI}\in\mathbb{C}^{N_I\times 1}$, $\mathbf{h}_{IT_B}\in\mathbb{C}^{N_I\times 1}$, $\mathbf{h}_{IT_U}\in\mathbb{C}^{N_I\times 1}$, and $\mathbf{h}_{R_BI}\in\mathbb{C}^{N_I\times 1}$, respectively, denote channel vectors from and to RIS, as explained after (\ref{eq:channel_blocks}) and illustrated in Fig. \ref{fig:channel}.
Similarly, the channel model in (\ref{eq:simplified_h2_noss}) which ignores the structural scattering writes as 
\begin{equation}\label{eq:simplified_h3_noss}
    \begin{aligned}
        \tilde{\mathbf{H}} &= \mathbf{H}_{RI}\mathbf{\Theta}\mathbf{H}_{IT}\\
        &= \left[\begin{matrix}
            \underbrace{\mathbf{h}_{R_BI}^\mathsf{T}\mathbf{\Theta}\mathbf{h}_{IT_B}}_{\text{loop-interference channel}} & \underbrace{\mathbf{h}_{R_BI}^\mathsf{T}\mathbf{\Theta}\mathbf{h}_{IT_U}}_{\text{user-RIS-base station channel}}\\
            \underbrace{\mathbf{h}_{R_DI}^\mathsf{T}\mathbf{\Theta}\mathbf{h}_{IT_B}}_{\text{base station-RIS-user channel}} & \underbrace{\mathbf{h}_{R_DI}^\mathsf{T}\mathbf{\Theta}\mathbf{h}_{IT_U}}_{\text{user-RIS-user channel}}
        \end{matrix}\right].
    \end{aligned}
\end{equation}
$\hfill\square$

We will next introduce the signal model and rate expression of the BD-RIS aided full-duplex system by considering the following two cases. 

\subsubsection{Structural Scattering}
By (\ref{eq:simplified_h3}), the received signals at the downlink user and at the base station are respectively 
\begin{subequations}
    \begin{align} 
        \non
        y_D = &\underbrace{\sqrt{P_D}\mathbf{h}_{R_DI}^\mathsf{T}(\mathbf{\Theta}-\mathbf{I}_{N_I})\mathbf{h}_{IT_B}s_D}_{\text{signal of interest at the downlink user}}\\
        &+ \underbrace{\sqrt{P_U}\mathbf{h}_{R_DI}^\mathsf{T}(\mathbf{\Theta}-\mathbf{I}_{N_I})\mathbf{h}_{IT_U}s_U}_{\text{interference from the uplink user}} + n_D, \\
        \non
        y_U = &\underbrace{\sqrt{P_U}\mathbf{h}_{R_BI}^\mathsf{T}(\mathbf{\Theta}-\mathbf{I}_{N_I})\mathbf{h}_{IT_U}s_U}_{\text{signal of interest at the base station}}\\ 
        &+ \underbrace{\sqrt{P_D}\mathbf{h}_{R_BI}^\mathsf{T}(\mathbf{\Theta}-\mathbf{I}_{N_I})\mathbf{h}_{IT_B}s_D}_{\text{interference from the base station}} + n_U,
 \end{align}
 \end{subequations}
where $P_D$ and $P_U$, respectively, denote the transmit power at the base station and the uplink user;  $s_D\in\mathbb{C}$ and $s_U\in\mathbb{C}$ with $\mathbb{E}\{|s_D|^2\} = 1$ and $\mathbb{E}\{|s_U|^2\}=1$, respectively, denote the independent transmit symbols at the base station and the uplink user; $n_D\sim\mathcal{CN}(0,\sigma_D^2)$ and $n_U\sim\mathcal{CN}(0,\sigma_U^2)$ denote the noise.

When the transmitting/receiving antenna is shared at the full-duplex base station, we have the channel reciprocity 
\begin{equation}
    \mathbf{h}_{IT_B} = \mathbf{h}_{R_BI} = \mathbf{h}_{BI}.
\end{equation} 
The downlink and uplink rates are thus  
\begin{subequations}
    \begin{align}
        \label{eq:rate_downlink}
        R_D &= \log_2\Big(1 + \frac{P_D|\mathbf{h}_{R_DI}^\mathsf{T}(\mathbf{\Theta}-\mathbf{I}_{N_I})\mathbf{h}_{BI}|^2}{P_U|\mathbf{h}_{R_DI}^\mathsf{T}(\mathbf{\Theta}-\mathbf{I}_{N_I})\mathbf{h}_{IT_U}|^2 + \sigma_D^2}\Big),\\
        \label{eq:rate_uplink}
        R_U &= \log_2\Big(1 + \frac{P_U|\mathbf{h}_{BI}^\mathsf{T}(\mathbf{\Theta}-\mathbf{I}_{N_I})\mathbf{h}_{IT_U}|^2}{P_D|\mathbf{h}_{BI}^\mathsf{T}(\mathbf{\Theta}-\mathbf{I}_{N_I})\mathbf{h}_{BI}|^2 + \sigma_U^2}\Big).
    \end{align}\label{eq:rate}
\end{subequations}

\subsubsection{No Structural Scattering} By (\ref{eq:simplified_h3_noss}), the received signals at the downlink user and the base station write as 
\begin{subequations}
\begin{align}
        \non
       \tilde{y}_D = &\underbrace{\sqrt{P_D}\mathbf{h}_{R_DI}^\mathsf{T}\mathbf{\Theta}\mathbf{h}_{IT_B}s_D}_{\text{signal of interest at the downlink user}}\\
        &~~~+ \underbrace{\sqrt{P_U}\mathbf{h}_{R_DI}^\mathsf{T}\mathbf{\Theta}\mathbf{h}_{IT_U}s_U}_{\text{interference from the uplink user}} + n_D, \\
        \non
        \tilde{y}_U = &\underbrace{\sqrt{P_U}\mathbf{h}_{R_BI}^\mathsf{T}\mathbf{\Theta}\mathbf{h}_{IT_U}s_U}_{\text{signal of interest at the base station}}\\
        &~~~+ \underbrace{\sqrt{P_D}\mathbf{h}_{R_BI}^\mathsf{T}\mathbf{\Theta}\mathbf{h}_{IT_B}s_D}_{\text{interference from the base station}} + n_U,
    \end{align}
\end{subequations}
which lead to the following rate expressions
\begin{subequations}
\begin{align}
        \label{eq:rate_downlink_noss}
        \tilde{R}_D &= \log_2\Big(1 + \frac{P_D|\mathbf{h}_{R_DI}^\mathsf{T}\mathbf{\Theta}\mathbf{h}_{BI}|^2}{P_U|\mathbf{h}_{R_DI}^\mathsf{T}\mathbf{\Theta}\mathbf{h}_{IT_U}|^2 + \sigma_D^2}\Big),\\
        \label{eq:rate_uplink_noss}
        \tilde{R}_U &= \log_2\Big(1 + \frac{P_U|\mathbf{h}_{BI}^\mathsf{T}\mathbf{\Theta}\mathbf{h}_{IT_U}|^2}{P_D|\mathbf{h}_{BI}^\mathsf{T}\mathbf{\Theta}\mathbf{h}_{BI}|^2 + \sigma_U^2}\Big).
    \end{align}\label{eq:rate_noss}
\end{subequations}

\subsection{Design Principle}

The rate expressions in (\ref{eq:rate}) and (\ref{eq:rate_noss}) have complex fractional terms, where the scattering matrix $\mathbf{\Theta}$ of BD-RIS appears in both the numerators and the denominators. This makes it difficult to study the effect of BD-RIS reciprocity in full-duplex systems. 
To ease the analysis, let us first focus on the received power of the signal of interest at the base station and the downlink user, i.e., the numerators of the fractional terms in (\ref{eq:rate}) and (\ref{eq:rate_noss}). 

\subsubsection{Structural Scattering}
In this case, the received power of the signal of interest at the downlink user is determined by the channel strength $\bar{P}_D = |\mathbf{h}_{R_DI}^\mathsf{T}(\mathbf{\Theta}-\mathbf{I}_{N_I})\mathbf{h}_{BI}|^2$, which is upper-bounded by 
\begin{equation}
        \label{eq:upperbound_downlink}
        \bar{P}_D^\mathsf{max} = (|\mathbf{h}_{R_DI}^\mathsf{T}\mathbf{h}_{BI}| + \|\mathbf{h}_{R_DI}\|_2\|\mathbf{h}_{BI}\|_2)^2,
\end{equation}
and the equality holds when 
\begin{equation}
        \label{eq:cond_downlink}
        \alpha_D \bar{\mathbf{h}}_{R_DI}^* = \mathbf{\Theta}\bar{\mathbf{h}}_{BI},
\end{equation}
where $\alpha_D = e^{\jmath\angle(-\mathbf{h}_{R_DI}^\mathsf{T}\mathbf{h}_{BI})}$, $\bar{\mathbf{h}}_{R_DI} = \mathbf{h}_{R_DI}\|\mathbf{h}_{R_DI}\|_2^{-1}$, and $\bar{\mathbf{h}}_{BI}=\mathbf{h}_{BI}\|\mathbf{h}_{BI}\|_2^{-1}$.
Similarly, the received power of the signal of interest at the base station is determined by the channel strength $\bar{P}_U = |\mathbf{h}_{BI}^\mathsf{T}(\mathbf{\Theta}-\mathbf{I}_{N_I})\mathbf{h}_{IT_U}|^2$, which is upper-bounded by 
\begin{equation}
        \label{eq:upperbound_uplink}
        \bar{P}_U^{\mathsf{max}} = (|\mathbf{h}_{IT_U}^\mathsf{T}\mathbf{h}_{BI}| + \|\mathbf{h}_{IT_U}\|_2\|\mathbf{h}_{BI}\|_2)^2,
\end{equation}
and the equality holds when 
\begin{equation}
        \label{eq:cond_uplink1}
        \alpha_U\bar{\mathbf{h}}_{BI}^* = \mathbf{\Theta}\bar{\mathbf{h}}_{IT_U}, 
\end{equation}
where $\alpha_U=e^{\jmath\angle(-\mathbf{h}_{BI}^\mathsf{T}\mathbf{h}_{IT_U})}$ and $\bar{\mathbf{h}}_{IT_U}=\mathbf{h}_{IT_U}\|\mathbf{h}_{IT_U}\|_2^{-1}$.

\subsubsection{No Structural Scattering}
In this case, we have $\tilde{P}_D = |\mathbf{h}_{R_DI}^\mathsf{T}\mathbf{\Theta}\mathbf{h}_{BI}|^2$ at the downlink user, upper-bounded by 
\begin{equation}
    \label{eq:upperbound_downlink_noss}
    \tilde{P}_D^\mathsf{max} = \|\mathbf{h}_{R_DI}\|_2^2\|\mathbf{h}_{BI}\|_2^2,
\end{equation}
where the equality holds when 
\begin{equation}
    \label{eq:cond_downlink_noss}
    \bar{\mathbf{h}}_{R_DI}^* = \mathbf{\Theta}\bar{\mathbf{h}}_{BI}.
\end{equation}
Similarly, at the base station we have $\tilde{P}_U = |\mathbf{h}_{BI}^\mathsf{T}\mathbf{\Theta}\mathbf{h}_{IT_U}|^2$, upper-bounded by 
\begin{equation}
    \label{eq:upperbound_uplink_noss}
    \tilde{P}_U^{\mathsf{max}} = \|\mathbf{h}_{IT_U}\|_2^2\|\mathbf{h}_{BI}\|_2^2,
\end{equation}
where the equality holds when 
\begin{equation}
        \label{eq:cond_uplink1_noss}
        \bar{\mathbf{h}}_{BI}^* = \mathbf{\Theta}\bar{\mathbf{h}}_{IT_U}.  
\end{equation}

In the following subsections, we will show in which conditions (\ref{eq:cond_downlink}) and (\ref{eq:cond_uplink1}), or (\ref{eq:cond_downlink_noss}) and (\ref{eq:cond_uplink1_noss}), can be simultaneously achieved by reciprocal and non-reciprocal BD-RISs.

\subsection{Reciprocal BD-RIS}

\subsubsection{Structural Scattering} 
In the case where the structural scattering is captured, we rewrite $\bar{P}_U$ as $\bar{P}_U = |\mathbf{h}_{IT_U}^\mathsf{T}(\mathbf{\Theta}^\mathsf{T}-\mathbf{I}_{N_I})\mathbf{h}_{BI}|^2$, such that having (\ref{eq:cond_uplink1}) to maximize $\bar{P}_U$ is equivalent to having 
\begin{equation}
    \label{eq:cond_uplink2}
    \alpha_U\bar{\mathbf{h}}_{IT_U}^* = \mathbf{\Theta}^\mathsf{T}\bar{\mathbf{h}}_{BI}.
\end{equation}

\subsubsection{No Structural Scattering}
Similarly, in the case with no structural scattering, we have $\tilde{P}_U = |\mathbf{h}_{IT_U}^\mathsf{T}\mathbf{\Theta}^\mathsf{T}\mathbf{h}_{BI}|^2$, such that having (\ref{eq:cond_uplink1_noss}) to maximize $\tilde{P}_U$ is equivalent to having 
\begin{equation}
    \label{eq:cond_uplink2_noss}
    \bar{\mathbf{h}}_{IT_U}^* = \mathbf{\Theta}^\mathsf{T}\bar{\mathbf{h}}_{BI}.
\end{equation}

Comparing (\ref{eq:cond_downlink}) and (\ref{eq:cond_uplink2}), or (\ref{eq:cond_downlink_noss}) and (\ref{eq:cond_uplink2_noss}), we observe that (\ref{eq:cond_downlink}) and (\ref{eq:cond_uplink2}) can be simultaneously achieved by the reciprocal BD-RIS with $\mathbf{\Theta} = \mathbf{\Theta}^\mathsf{T}$ if and only if $\alpha_D\bar{\mathbf{h}}_{R_DI}^*=\alpha_U\bar{\mathbf{h}}_{IT_U}^*$, or (\ref{eq:cond_downlink_noss}) and (\ref{eq:cond_uplink2_noss}) can be simultaneously achieved by the reciprocal BD-RIS if and only if $\bar{\mathbf{h}}_{R_DI}^*=\bar{\mathbf{h}}_{IT_U}^*$. This corresponds to a very specific case where the uplink and downlink users are aligned and share the same small-scale fading. 
In other words, with generally $\alpha_D\bar{\mathbf{h}}_{R_DI}^*\ne\alpha_U\bar{\mathbf{h}}_{IT_U}^*$, or $\bar{\mathbf{h}}_{R_DI}^*\ne\bar{\mathbf{h}}_{IT_U}^*$ in wireless communication systems, the maximum received power of the signal of interest at the downlink user and at the base station cannot be simultaneously obtained by the reciprocal BD-RIS. 

To facilitate the analysis, when reciprocal BD-RIS is adopted, we will choose to always achieve the performance upper-bound of the channel strength at the uplink, i.e., $\bar{P}_U^\mathsf{max}$ in (\ref{eq:upperbound_uplink}) or $\tilde{P}_U^\mathsf{max}$ in (\ref{eq:upperbound_uplink_noss}), based on different locations of the uplink user\footnote{In this work, although we focus on the uplink when applying reciprocal BD-RIS, the derivations also hold for the case of focusing on the downlink. The observations will not change except that uplink and downlink are shuffled.}. 
With the RIS design focusing purely on the uplink at hand, we will further check the performance at the downlink. Irrespective of the channel fading conditions, this can be generally done by applying the closed-form solution for reciprocal BD-RIS proposed in \cite{nerini2023closed}.
More detailed reciprocal BD-RIS designs will be explained in the performance evaluation section.

\subsection{Non-Reciprocal BD-RIS}
In this case, it is also difficult to find a closed-form expression of $\mathbf{\Theta}$ to exactly achieve  (\ref{eq:cond_downlink}) and (\ref{eq:cond_uplink1}), or (\ref{eq:cond_downlink_noss}) and (\ref{eq:cond_uplink1_noss}).
To facilitate the analysis, we turn to find such an approximated solution by solving the following matrix projection problem:
\begin{equation}
    \label{eq:assym_opt}
    \mathbf{\Theta}^{\mathsf{NR}} = \mathop{\arg\min}\limits_{\mathbf{\Theta}^\mathsf{H}\mathbf{\Theta}=\mathbf{I}_{N_I}}\|\mathbf{X} - \mathbf{\Theta}\mathbf{Y}\|_\mathsf{F}^2,
\end{equation}
where 
\begin{subequations}
    \begin{align}
        \mathbf{X} &= \begin{cases}[\alpha_D\bar{\mathbf{h}}_{R_DI}^*,\alpha_U\bar{\mathbf{h}}_{BI}^*], &\text{structural scattering},\\ [\bar{\mathbf{h}}_{R_DI}^*,\bar{\mathbf{h}}_{BI}^*], &\text{no structural scattering},
        \end{cases}\\
        \mathbf{Y} &= [\bar{\mathbf{h}}_{BI},\bar{\mathbf{h}}_{IT_U}]. 
    \end{align}
\end{subequations}
Then we have the following proposition. 

\textit{Proposition 1:} Problem  (\ref{eq:assym_opt}) has the following properties:

\textit{Property 1:} Given the singular value decomposition (SVD) $\mathbf{X}\mathbf{Y}^\mathsf{H} = \mathbf{U}\mathbf{\Sigma}\mathbf{V}^\mathsf{H}$ with unitary matrices $\mathbf{U}\in\mathbb{C}^{N_I\times N_I}$ and $\mathbf{V}\in\mathbb{C}^{N_I\times N_I}$, there exists a globally optimal solution 
\begin{equation}
    \mathbf{\Theta}^{\mathsf{NR}} = \mathbf{U}\mathbf{V}^\mathsf{H}.\label{eq:opt_assym_theta}
\end{equation}

\textit{Property 2:} Problem  (\ref{eq:assym_opt}) has a global minimum as 
\begin{equation}
    \|\mathbf{X} - \mathbf{\Theta}^{\mathsf{NR}}\mathbf{Y}\|_\mathsf{F}^2 = 4 - 2\mathsf{Tr}(\mathbf{\Sigma}).\label{eq:minimum}
\end{equation} 

\textit{Property 3:} $\mathsf{Tr}(\mathbf{\Sigma})$ is upper-bounded by 
\begin{equation}
    \mathsf{Tr}(\mathbf{\Sigma}) \overset{\text{(b)}}{\le} \mathsf{Tr}(\mathbf{\Sigma}_\mathbf{X}\mathbf{\Sigma}_\mathbf{Y})=\sigma_{\mathbf{X},1}\sigma_{\mathbf{Y},1} + \sigma_{\mathbf{X},2}\sigma_{\mathbf{Y},2},\label{eq:bound1}
\end{equation} 
where $\mathbf{\Sigma}_{\mathbf{X}}=[\mathsf{diag}(\sigma_{\mathbf{X},1},\sigma_{\mathbf{X},1});\mathbf{0}_{(N_I-2)\times 2}]\in\mathbb{R}^{N_I\times 2}$ and $\mathbf{\Sigma}_{\mathbf{Y}}=[\mathsf{diag}(\sigma_{\mathbf{Y},1},\sigma_{\mathbf{Y},1}),\mathbf{0}_{2\times(N_I-2)}]\in\mathbb{R}^{2\times N_I}$, respectively, contain the singular values of $\mathbf{X}$ and $\mathbf{Y}^\mathsf{H}$ in descending order. Specifically, the equality of (b) holds when $\mathbf{X}$ and $\mathbf{Y}^\mathsf{H}$ have aligned singular vectors. 

\textit{Property 4:} $\sigma_{\mathbf{X},1}\sigma_{\mathbf{Y},1} + \sigma_{\mathbf{X},2}\sigma_{\mathbf{Y},2}$ is bounded by
\begin{equation}
    \sqrt{2}\overset{\text{(c)}}{\le} \sigma_{\mathbf{X},1}\sigma_{\mathbf{Y},1} + \sigma_{\mathbf{X},2}\sigma_{\mathbf{Y},2} \overset{\text{(d)}}{\le} 2,\label{eq:bound2}
\end{equation}
where the equality of (c) holds when $\sigma_{j,1} = 1$ and $\sigma_{j',1} = \sqrt{2}$, $\forall j\ne j',j,j'\in\{\mathbf{X},\mathbf{Y}\}$; the equality of (d) holds when $\sigma_{\mathbf{X},1} = \sigma_{\mathbf{Y},1}$ and $\sigma_{\mathbf{X},2}=\sigma_{\mathbf{Y},2}$. 

\textit{Proof:} Please refer to the Appendix. $\hfill\square$

Based on Proposition 1, we have the following corollary. 

\textit{Corollary 1:} When $\mathbf{X}$ and $\mathbf{Y}^\mathsf{H}$ have aligned singular vectors and identical singular values, the conditions (\ref{eq:cond_downlink}) and (\ref{eq:cond_uplink1}), or the conditions (\ref{eq:cond_downlink_noss}) and (\ref{eq:cond_uplink1_noss}), can be simultaneously achieved by $\mathbf{\Theta}^\mathsf{NR}$ in (\ref{eq:opt_assym_theta}). 

\textit{Proof:} With $\mathbf{X}$ and $\mathbf{Y}^\mathsf{H}$ satisfying the conditions to simultaneously achieve equalities (b) and (d), we have $\mathsf{Tr}(\mathbf{\Sigma}) = 2$. This implies $\|\mathbf{X}-\mathbf{\Theta}^{\mathsf{NR}}\mathbf{Y}\|_\mathsf{F} = 0$, or, equivalently, (\ref{eq:cond_downlink}) and (\ref{eq:cond_uplink1}), or (\ref{eq:cond_downlink_noss}) and (\ref{eq:cond_uplink1_noss}), hold true simultaneously, which completes the proof. $\hfill\square$

\textit{Remark 3:} 
When conditions (\ref{eq:cond_downlink}) and (\ref{eq:cond_uplink1}) are simultaneously achieved by reciprocal BD-RIS, we have $\mathbf{X} = \alpha_U[\bar{\mathbf{h}}_{IT_U}^*,\bar{\mathbf{h}}_{BI}^*] = \alpha_U\mathbf{Y}^*\mathbf{\Gamma}$ with $\mathbf{\Gamma} = [0~1;1~0]$; similarly, when conditions (\ref{eq:cond_downlink_noss}) and (\ref{eq:cond_uplink1_noss}) are simultaneously achieved by reciprocal BD-RIS, we have $\mathbf{X} = [\bar{\mathbf{h}}_{IT_U}^*,\bar{\mathbf{h}}_{BI}^*] = \mathbf{Y}^*\mathbf{\Gamma}$, both of which are special cases of having $\mathbf{X}$ and $\mathbf{Y}^\mathsf{H}$ with identical singular values and aligned singular vectors.
This implies that the condition to simultaneously achieve (\ref{eq:cond_downlink}) and (\ref{eq:cond_uplink1}), or to simultaneously achieve (\ref{eq:cond_downlink_noss}) and (\ref{eq:cond_uplink1_noss}), by non-reciprocal BD-RIS, i.e., Corollary 1, is much less strict than that by reciprocal BD-RIS. This statement will be numerically shown in the following section.

\section{Performance Evaluation}
\label{sec:simulation}

To visualize the effect of BD-RIS reciprocity in full-duplex systems, we assume a uniform linear array at the BD-RIS with half-wavelength inter-element spacing, and line of sight propagation between the base station, BD-RIS, and users. 
As illustrated in Fig. \ref{fig:case}, we assume the transmit signal from the base station impinges on the BD-RIS with angle $\phi_{BI}\in[0,\pi]$ and is reflected towards the downlink user with angle $\phi_{R_DI}\in[0,\pi]$; the transmit signal from the uplink user impinges on the BD-RIS with angle $\phi_{IT_U}\in[0,\pi]$ and is reflected towards the base station with angle $\phi_{BI}$. 
Therefore, the line of sight channels are modeled as $\mathbf{h}_o = \sqrt{\zeta_o}\bar{\mathbf{h}}_o$, $\forall o\in\{BI,R_DI,IT_U\}$, where $\zeta_o$ denotes the large-scale fading  and 
\begin{equation}
    \bar{\mathbf{h}}_o=\frac{1}{\sqrt{N_I}}[1,e^{\jmath\pi\cos\phi_o},\ldots,e^{\jmath\pi(N_I-1)\cos\phi_o}]^\mathsf{T},
\end{equation} 
denotes the small-scale fading.
Specifically, the large-scale fading is modeled as $\zeta_o = \zeta_0(\frac{d_o}{d_0})^{-\epsilon}$ with $\zeta_0$ the signal attenuation at a reference distance $d_0$, $d_o$ the distance between devices, and $\epsilon$ the pass loss exponent.

Recall that when reciprocal BD-RIS is adopted, we choose to always guarantee $\bar{P}_U^\mathsf{max}$ or $\tilde{P}_U^\mathsf{max}$ is achieved based on locations of the uplink user, and check how the downlink performance will be like. In line of sight propagation, 
different reciprocal BD-RIS architectures achieve the same channel strength upper-bound \cite{shen2021,nerini2023closed}. In other words, in line of sight propagation, D-RIS is sufficient to achieve $\bar{P}_U^\mathsf{max}$ or $\tilde{P}_U^\mathsf{max}$ with a closed-form solution.
Specifically, $\bar{P}_U^\mathsf{max}$ capturing the structural scattering can be achieved by the following closed-form optimal solution
\begin{equation}\label{eq:opt_sym_theta}
    \begin{aligned}
        \mathbf{\Theta}^\mathsf{R} &= \mathsf{diag}(e^{\jmath\theta_1},\ldots,e^{\jmath\theta_{N_I}}), \\
        \theta_n &= 
            \angle\alpha_U - \pi(n-1)(\cos\phi_{BI} + \cos\phi_{IT_U}), \\
             &~~~~~~~~~~~~~~~~~~~~~~~~~~~~~~~~~\forall n = 1,\ldots,N_I,
    \end{aligned} 
\end{equation}
such that the upper-bound (\ref{eq:upperbound_uplink}) writes as 
\begin{equation}
    \begin{aligned}
        \bar{P}_U^\mathsf{max} &= \zeta_{BI}\zeta_{IT_U}\Big(\frac{1}{N_I}\Big|\sum_{n=1}^{N_I}e^{\jmath\pi (n-1)(\cos\phi_{BI} + \cos\phi_{IT_U})}\Big|+1\Big)^2\\
        \label{eq:upperbound_los}
        &\overset{\text{(e)}}{\le}4\zeta_{BI}\zeta_{IT_U},
    \end{aligned}
\end{equation}
where the equality of (e) holds true when $\phi_{BI} + \phi_{IT_U} = \pi$, which corresponds to the specular reflection such that the RIS works as a mirror.
Similarly, when the structural scattering is ignored, $\tilde{P}_U^\mathsf{max}$ can be achieved by the following closed-form optimal solution:
\begin{equation}\label{eq:opt_sym_theta_noss}
    \begin{aligned}
        \tilde{\mathbf{\Theta}}^\mathsf{R} &= \mathsf{diag}(e^{\jmath\tilde{\theta}_1},\ldots,e^{\jmath\tilde{\theta}_{N_I}}), \\
        \tilde{\theta}_n &= 
             - \pi(n-1)(\cos\phi_{BI} + \cos\phi_{IT_U}), \forall n = 1,\ldots,N_I,
    \end{aligned} 
\end{equation}
such that the upper-bound (\ref{eq:upperbound_uplink_noss}) writes as 
\begin{equation}
    \tilde{P}_U^\mathsf{max} = \zeta_{BI}\zeta_{IT_U}.
    \label{eq:upperbound_los_noss}
\end{equation}

\textit{Remark 4:} 
Comparing (\ref{eq:upperbound_los}) with (\ref{eq:upperbound_los_noss}), we observe that including structural scattering in the channel model leads to up to four times of channel gain in line of sight propagation. This is because, in addition to tuning the $\mathbf{\Theta}$ of RIS, the structural scattering provides an additional degree of freedom, that is tuning the locations of the transmitter and receiver, to further boost the channel strength. Specifically, the maximum channel strength in (\ref{eq:upperbound_los}) is achieved when $\mathbf{\Theta}$ is properly designed and the base station and the uplink user is symmetrically located based on the RIS. The latter condition can only be observed when the structural scattering is captured. This is also the primary reason that (\ref{eq:upperbound_los}) can be up to four times of (\ref{eq:upperbound_los_noss}).

The above closed-form solutions, $\mathbf{\Theta}^{\mathsf{R}}$ and $\tilde{\mathbf{\Theta}}^{\mathsf{R}}$, will be used to analyze the case for reciprocal BD-RIS, while the closed-form solution given in Proposition 1 will be used to analyze the case for non-reciprocal BD-RIS. 
Unless otherwise stated, we set $\zeta_0 = -30$ dB, $d_0 = 1$ m, $\epsilon = 2$, $d_{BI} = 30$ m, $d_{R_DI} = d_{IT_U} = 5$ m, $\sigma_D^2 = \sigma_U^2 = -80$ dBm, and $P_D = P_U = 500$ mW.
In the following simulations, we will show the effect of BD-RIS reciprocity in full-duplex systems in beam pattern, channel strength, and rate, each of which will capture the impact of structural scattering.

\subsection{Beam Pattern}

\begin{figure}[t]
    \centering
    \includegraphics[width = 0.485\textwidth]{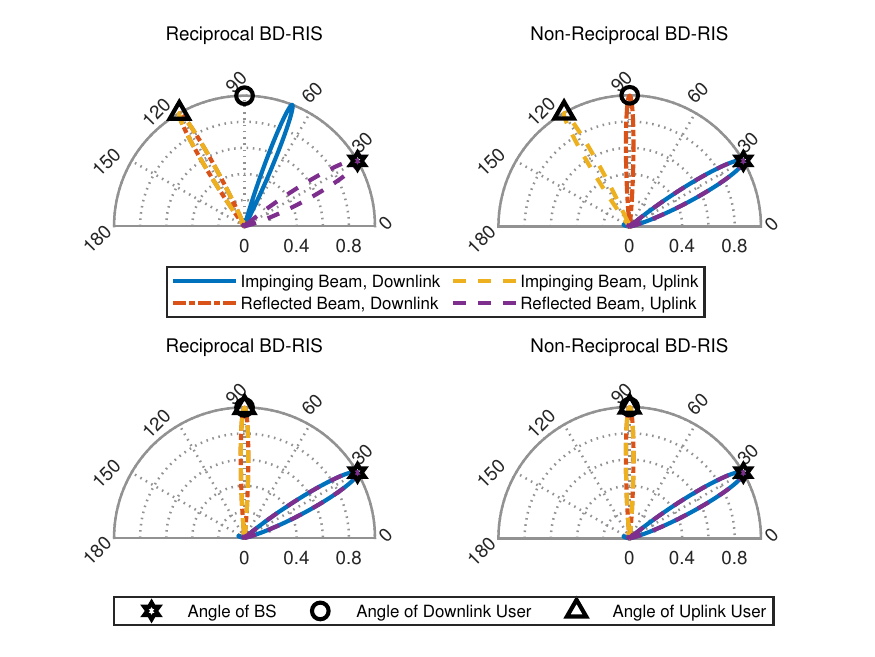}
    \caption{The impinging and reflected beam patterns of reciprocal and non-reciprocal BD-RISs without structural scattering ($\phi_{BI} = \frac{\pi}{6}$, $\phi_{R_DI} = \frac{\pi}{2}$, $N_I=16$). Top: $\phi_{IT_U} = \frac{2\pi}{3}$; bottom: $\phi_{IT_U} = \frac{\pi}{2}$.}
    \label{fig:beam_noss}
\end{figure}

\subsubsection{No Structural Scattering} 
Fig. \ref{fig:beam_noss} plots the impinging and reflected beam patterns of reciprocal and non-reciprocal BD-RISs (normalized by the channel strength upper-bound $\tilde{P}_U^\mathsf{max}$ in (\ref{eq:upperbound_los_noss})), each of which is defined as a function of $\phi\in[0,\pi]$:
\begin{subequations}
    \begin{align} 
        &\tilde{P}_D^\mathsf{impinging}(\phi) = |\bar{\mathbf{h}}_{R_DI}^\mathsf{T}\mathbf{\Theta}\bar{\mathbf{a}}(\phi)|^2,\\  &\tilde{P}_D^\mathsf{reflected}(\phi) = |\bar{\mathbf{a}}^\mathsf{T}(\phi)\mathbf{\Theta}\bar{\mathbf{h}}_{BI}|^2,\\
        &\tilde{P}_U^\mathsf{impinging}(\phi) = |\bar{\mathbf{h}}_{BI}^\mathsf{T}\mathbf{\Theta}\bar{\mathbf{a}}(\phi)|^2,\\
        &\tilde{P}_U^\mathsf{reflected}(\phi) = |\bar{\mathbf{a}}^\mathsf{T}(\phi)\mathbf{\Theta}\bar{\mathbf{h}}_{IT_U}|^2,
    \end{align}
\end{subequations}
where $\bar{\mathbf{a}}(\phi) = \frac{1}{\sqrt{N_I}}[1,e^{\jmath\pi\cos\phi},\ldots,e^{\jmath\pi(N_I-1)\cos\phi}]^\mathsf{T}\in\mathbb{C}^{N_I\times 1}$ is the steering vector, $\mathbf{\mathbf{\Theta}} = \tilde{\mathbf{\Theta}}^\mathsf{R}$ in (\ref{eq:opt_sym_theta_noss}) for reciprocal BD-RIS, and $\mathbf{\Theta} = \mathbf{\Theta}^\mathsf{NR}$ in (\ref{eq:opt_assym_theta}) for non-reciprocal BD-RISs.
The BS in Fig. \ref{fig:beam_noss} and the following figures refers to the base station. 
In the top two figures, we choose $\phi_{BI} = \frac{\pi}{6}$, $\phi_{R_DI} = \frac{\pi}{2}$, and $\phi_{IT_U} = \frac{2\pi}{3}$, which mathematically leads to $\mathbf{X}$ and $\mathbf{Y}^\mathsf{H}$ with nearly aligned singular vectors and similar singular values. Simulation results show that both the impinging and reflected beams of non-reciprocal BD-RIS can exactly point to the directions of interest, while those of reciprocal BD-RIS fail to point to the downlink user. 
For example, the blue beam for reciprocal BD-RIS does not point to the base station located at $\phi_{BI} = \frac{\pi}{6}$, and the red beam for reciprocal BD-RIS fails to point to the downlink user located at $\phi_{R_DI} = \frac{\pi}{2}$.
This verifies Corollary 1 and shows the benefit of using non-reciprocal BD-RIS in full-duplex systems.
In the bottom two figures, we choose $\phi_{BI} = \frac{\pi}{6}$, $\phi_{R_DI} = \frac{\pi}{2}$, and $\phi_{IT_U} = \frac{\pi}{2}$ (hence uplink and downlink users are aligned), which mathematically leads to $\mathbf{X} = \mathbf{Y}^*\mathbf{\Gamma}$. Simulation results show that both the impinging and reflected beams of reciprocal and non-reciprocal BD-RISs can exactly point to the directions of interest, which verify the discussion in Remark 3. 

\begin{figure}[t]
    \centering
    \includegraphics[width = 0.485\textwidth]{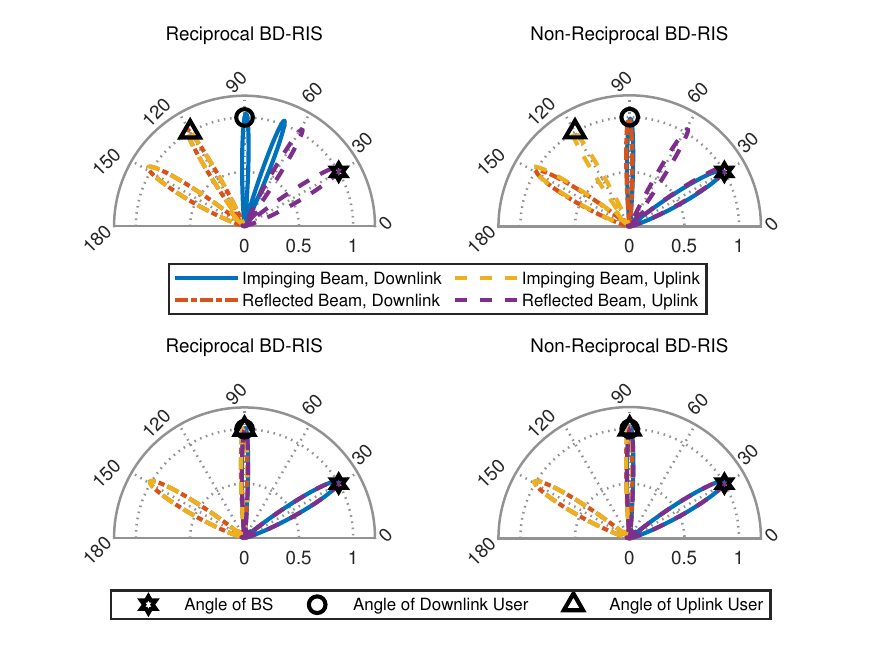}
    \caption{The impinging and reflected beam patterns of reciprocal and non-reciprocal BD-RISs with structural scattering ($\phi_{BI} = \frac{\pi}{6}$, $\phi_{R_DI} = \frac{\pi}{2}$, $N_I=16$). Top: $\phi_{IT_U} = \frac{2\pi}{3}$; bottom: $\phi_{IT_U} = \frac{\pi}{2}$.}
    \label{fig:beam}
\end{figure}

\subsubsection{Structural Scattering}
Fig. \ref{fig:beam} plots the impinging and reflected beam patterns (normalized by the upper-bound (\ref{eq:upperbound_los})) of reciprocal and non-reciprocal BD-RISs, each of which is defined as
\begin{subequations}
    \begin{align} 
        &\bar{P}_D^\mathsf{impinging}(\phi) = \frac{|\bar{\mathbf{h}}_{R_DI}^\mathsf{T}(\mathbf{\Theta} - \mathbf{I}_{N_I})\bar{\mathbf{a}}(\phi)|^2}{(|\bar{\mathbf{h}}_{IT_U}^\mathsf{T}\bar{\mathbf{h}}_{BI}|+1)^2},\\  &\bar{P}_D^\mathsf{reflected}(\phi) = \frac{|\bar{\mathbf{a}}^\mathsf{T}(\phi)(\mathbf{\Theta} - \mathbf{I}_{N_I})\bar{\mathbf{h}}_{BI}|^2}{(|\bar{\mathbf{h}}_{IT_U}^\mathsf{T}\bar{\mathbf{h}}_{BI}|+1)^2},\\
        &\bar{P}_U^\mathsf{impinging}(\phi) = \frac{|\bar{\mathbf{h}}_{BI}^\mathsf{T}(\mathbf{\Theta} - \mathbf{I}_{N_I})\bar{\mathbf{a}}(\phi)|^2}{(|\bar{\mathbf{h}}_{IT_U}^\mathsf{T}\bar{\mathbf{h}}_{BI}|+1)^2},\\
        &\bar{P}_U^\mathsf{reflected}(\phi) = \frac{|\bar{\mathbf{a}}^\mathsf{T}(\phi)(\mathbf{\Theta} - \mathbf{I}_{N_I})\bar{\mathbf{h}}_{IT_U}|^2}{(|\bar{\mathbf{h}}_{IT_U}^\mathsf{T}\bar{\mathbf{h}}_{BI}|+1)^2},
    \end{align}
\end{subequations}
where $\mathbf{\mathbf{\Theta}} = \tilde{\mathbf{\Theta}}^\mathsf{R}$ in (\ref{eq:opt_sym_theta}) for reciprocal BD-RIS, and $\mathbf{\Theta} = \mathbf{\Theta}^\mathsf{NR}$ in (\ref{eq:opt_assym_theta}) for non-reciprocal BD-RISs.
Similar conclusions as in Fig. \ref{fig:beam_noss} can be obtained. 
In addition, comparing Fig. \ref{fig:beam} with Fig. \ref{fig:beam_noss}, we find additional impinging and reflected beams in Fig. \ref{fig:beam} which do not point to the directions of interest. This essentially comes from the structural scattering at BD-RIS, which implies that when accurately capturing the structural scattering, additional users can be potentially served without sacrificing the performance of the original two users.

\subsection{Channel Strength}

\subsubsection{No Structural Scattering}
\begin{figure}[t]
    \centering
   \includegraphics[width = 0.48\textwidth]{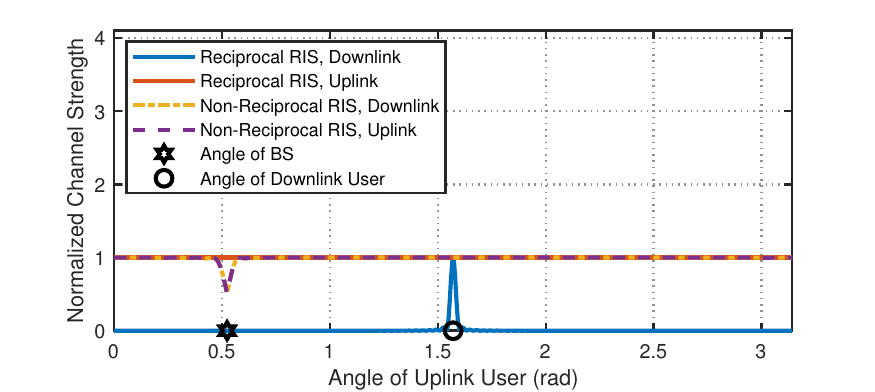}
    \caption{Normalized channel strength versus the location of the uplink user using reciprocal and non-reciprocal BD-RISs without structural scattering ($\phi_{BI} = \frac{\pi}{6}$, $\phi_{R_DI} = \frac{\pi}{2}$, $N_I = 64$).}
    \label{fig:channel_strength_noss}
\end{figure}

Fig. \ref{fig:channel_strength_noss} illustrates the normalized channel strength 
\begin{subequations}
    \begin{align}
        &\bar{\tilde{P}}_U = |\bar{\mathbf{h}}_{BI}^\mathsf{T}\mathbf{\Theta}\bar{\mathbf{h}}_{IT_U}|^2,\\
        &\bar{\tilde{P}}_D = |\bar{\mathbf{h}}_{R_DI}^\mathsf{T}\mathbf{\Theta}\bar{\mathbf{h}}_{BI}|^2,
    \end{align}
\end{subequations} 
versus the location of the uplink user using reciprocal BD-RIS with $\mathbf{\mathbf{\Theta}} = \tilde{\mathbf{\Theta}}^\mathsf{R}$ in (\ref{eq:opt_sym_theta_noss}), and non-reciprocal BD-RISs with $\mathbf{\Theta} = \mathbf{\Theta}^\mathsf{NR}$ in (\ref{eq:opt_assym_theta}) obtained by $\mathbf{X}$ with no structural scattering. 
We have the following observations. 
\textit{First}, when the uplink user is aligned with the base station, i.e., $\phi_{IT_U} = \phi_{BI}$, the non-reciprocal BD-RIS achieves the worst performance in both uplink and downlink. 
In this special case, two transmitters are aligned such that only one narrow impinging beam is needed, while two receivers are located far from each other such that two separate reflected beams are needed. 
The different requirements for impinging beams and reflected beams cannot be simultaneously and optimally enabled by non-reciprocal BD-RIS. 
Mathematically, this is because this deployment leads to $\sigma_{\mathbf{Y},1} = \sqrt{2}$ and $\sigma_{\mathbf{X},1}\approx\sigma_{\mathbf{X},2}\approx 1$, which corresponds to the lower bound in Property 4 of Proposition 1. In this case, the solution $\mathbf{\Theta}^{\mathsf{NR}}$ results in a large global minimum $\|\mathbf{X}-\mathbf{\Theta}^{\mathsf{NR}}\mathbf{Y}\|_\mathsf{F}^2$ and thus a reduced channel strength performance. 
\textit{Second}, when the uplink user is aligned with the downlink user, i.e., $\phi_{R_DI} = \phi_{IT_U}$, both non-reciprocal and reciprocal BD-RISs achieve the same performance following the interpretation given in Fig. \ref{fig:beam_noss}.
\textit{Third}, when the uplink user is in other locations, the reciprocal BD-RIS cannot simultaneously serve two links, while the non-reciprocal BD-RIS can. This validates the discussion in Remark 3.

\subsubsection{Structural Scattering}
\begin{figure}[t]
    \centering
    \subfigure[][$\phi_{R_DI} = \frac{\pi}{2}$]{\includegraphics[width = 0.48\textwidth]{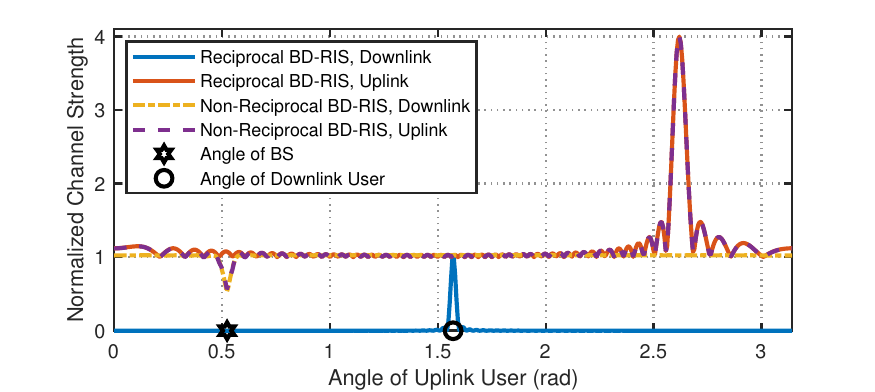}}
    \subfigure[][$\phi_{R_DI} = \frac{5\pi}{6}$]{\includegraphics[width = 0.48\textwidth]{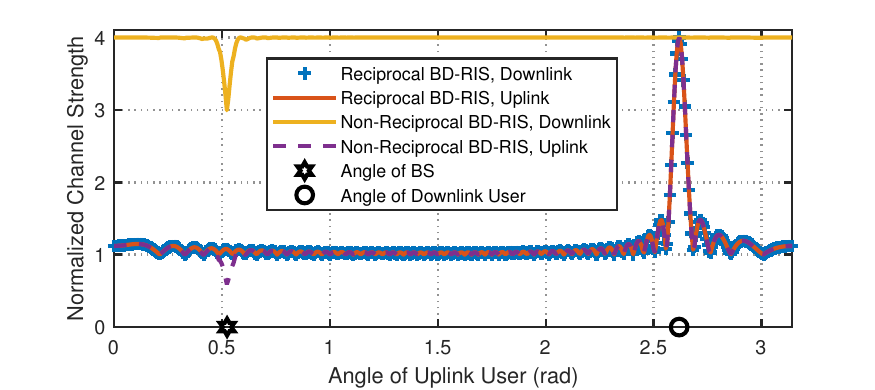}}
    \caption{Normalized channel strength versus the location of the uplink user using reciprocal and non-reciprocal BD-RISs with structural scattering ($\phi_{BI} = \frac{\pi}{6}$, $N_I = 64$).}
    \label{fig:channel_strength}
\end{figure}

Fig. \ref{fig:channel_strength} illustrates the normalized channel strength
\begin{subequations}\label{eq:normalized_cs}
    \begin{align}
        &\bar{\bar{P}}_U = |\bar{\mathbf{h}}_{BI}^\mathsf{T}(\mathbf{\Theta} - \mathbf{I}_{N_I})\bar{\mathbf{h}}_{IT_U}|^2,\\ 
        &\bar{\bar{P}}_D = |\bar{\mathbf{h}}_{R_DI}^\mathsf{T}(\mathbf{\Theta} - \mathbf{I}_{N_I})\bar{\mathbf{h}}_{BI}|^2, 
    \end{align}
\end{subequations}
versus the location of the uplink user using reciprocal BD-RIS with $\mathbf{\Theta} = \mathbf{\Theta}^\mathsf{R}$ in (\ref{eq:opt_sym_theta}) and non-reciprocal BD-RISs with $\mathbf{\Theta} = \mathbf{\Theta}^\mathsf{NR}$ in (\ref{eq:opt_assym_theta}) obtained by $\mathbf{X}$ with structural scattering. 
In addition to the conclusions obtained from Fig. \ref{fig:channel_strength_noss}, we have the following observations.
\textit{First}, as shown in Fig. \ref{fig:channel_strength}(a), when the uplink user is located at $\phi_{IT_U} = \pi-\phi_{BI}$, both reciprocal and non-reciprocal BD-RISs have the best performance in the uplink since the maximum in (\ref{eq:upperbound_los}) is achieved. This corresponds to the case where the uplink user and base station are symmetrically located and the BD-RIS works as a mirror. 
\textit{Second}, as shown in Fig. \ref{fig:channel_strength}(b), when the downlink user is located at $\phi_{R_DI} = \pi-\phi_{BI}$, the channel strength in the downlink applying the reciprocal BD-RIS can be as much as that in the uplink benefiting from the structural scattering. More importantly, when the downlink and uplink users are aligned, and are symmetrically located with base station, both reciprocal and non-reciprocal BD-RIS achieve maximum channel strength in the uplink and downlink. This validates the discussions in Remarks 3 and 4.
\textit{Third}, comparing Fig. \ref{fig:channel_strength} with Fig. \ref{fig:channel_strength_noss}, the normalized channel strength taking into account to the structural scattering could be up to four times of that ignoring the structural scattering. 
Specifically, we can observe that $\bar{\bar{P}}_U = 4$ at $\phi_{IT_U} = \pi-\phi_{BI}$, while  $\bar{\tilde{P}}_U = 1$ at the same point.
This verifies the theoretical upper-bounds given in (\ref{eq:upperbound_los}) and (\ref{eq:upperbound_los_noss}), validates the discussion in Remark 4, and thus demonstrates the benefit of accurately capturing structural scattering in enhancing channel strength.

\begin{figure}[t]
    \centering
    \subfigure[][$\phi_{R_DI} = \frac{\pi}{2}$]{\includegraphics[width = 0.48\textwidth]{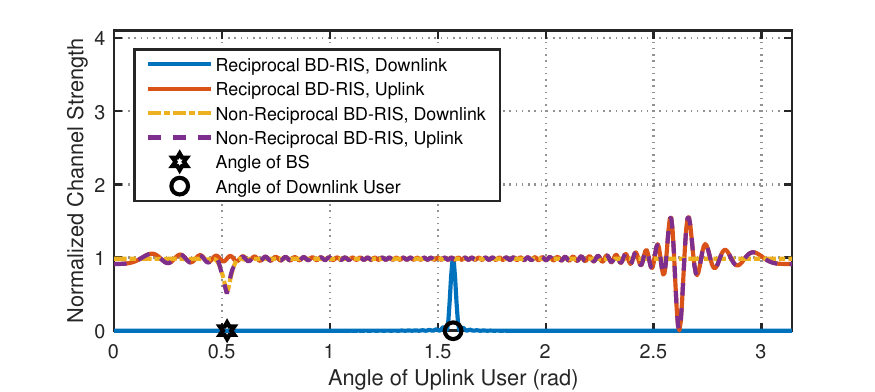}}
    \subfigure[][$\phi_{R_DI} = \frac{5\pi}{6}$]{\includegraphics[width = 0.48\textwidth]{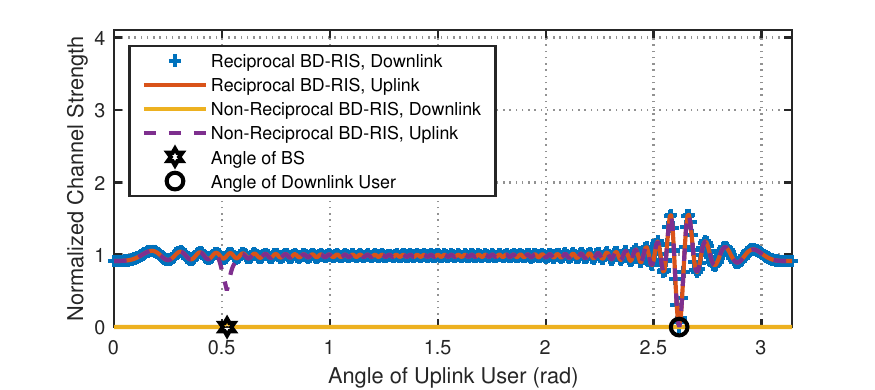}}
    \caption{Normalized channel strength versus the location of the uplink user using reciprocal and non-reciprocal BD-RISs without structural scattering for design and with for evaluation ($\phi_{BI} = \frac{\pi}{6}$, $N_I = 64$).}
    \label{fig:channel_strength_deswoss_evawss}
\end{figure}

\subsubsection{Design without and Evaluation with Structural Scattering}
Fig. \ref{fig:channel_strength_deswoss_evawss} illustrates the normalized channel strength in (\ref{eq:normalized_cs}) versus the location of the uplink user using reciprocal BD-RIS with $\mathbf{\Theta} = \tilde{\mathbf{\Theta}}^\mathsf{R}$ in (\ref{eq:opt_sym_theta_noss}) and $\mathbf{\Theta} = \mathbf{\Theta}^\mathsf{NR}$ in (\ref{eq:opt_assym_theta}) obtained by $\mathbf{X}$ with no structural scattering. Comparing Fig. \ref{fig:channel_strength_deswoss_evawss} with Fig. \ref{fig:channel_strength}, we observe that the benefit from structural scattering in both uplink and downlink disappears due to the ignorance of structural scattering for BD-RIS design. More importantly, when the uplink (downlink) user is symmetrically located with the base station, the normalized channel strength becomes zero. This can be mathematically explained by plugging $\tilde{\mathbf{\Theta}}^\mathsf{R}$ into (\ref{eq:normalized_cs}a), which yields
\begin{equation}
    \begin{aligned}
        \bar{\bar{P}}_U &= |\bar{\mathbf{h}}_{BI}^\mathsf{T}(\tilde{\mathbf{\Theta}}^\mathsf{R} - \mathbf{I}_{N_I})\bar{\mathbf{h}}_{IT_U}|^2\\
        &= \Big(1-\frac{1}{N_I}\Big(\sum_{n=1}^{N_I}e^{\jmath\pi (n-1)(\cos\phi_{BI} + \cos\phi_{IT_U})}\Big)\Big)^2\\
        &\overset{\text{(f)}}{=} 0,
    \end{aligned}
\end{equation}
where (f) holds since $\phi_{BI} = \pi - \phi_{IT_U}$. This demonstrates that, with specific locations of users and base station, designing BD-RIS without capturing the structural scattering will cancel the benefit from tuning $\mathbf{\Theta}$ in real communication systems.

\begin{figure}[t]
    \centering
    \includegraphics[width = 0.48\textwidth]{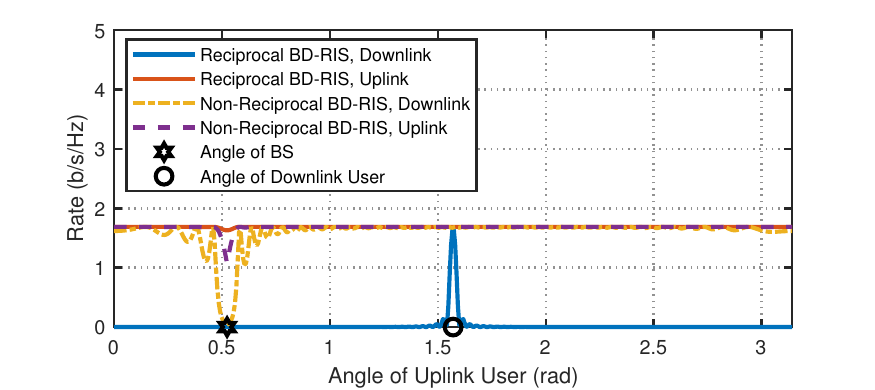}
    \caption{The uplink and downlink rates versus the location of the uplink user using reciprocal and non-reciprocal BD-RISs without structural scattering ($\phi_{BI} = \frac{\pi}{6}$, $\phi_{R_DI} = \frac{\pi}{2}$, $N_I=64$).}
    \label{fig:rate_noss}
\end{figure}

\begin{figure}[t]
    \centering
    \subfigure[][$\phi_{R_DI} = \frac{\pi}{2}$]{
    \includegraphics[width = 0.48\textwidth]{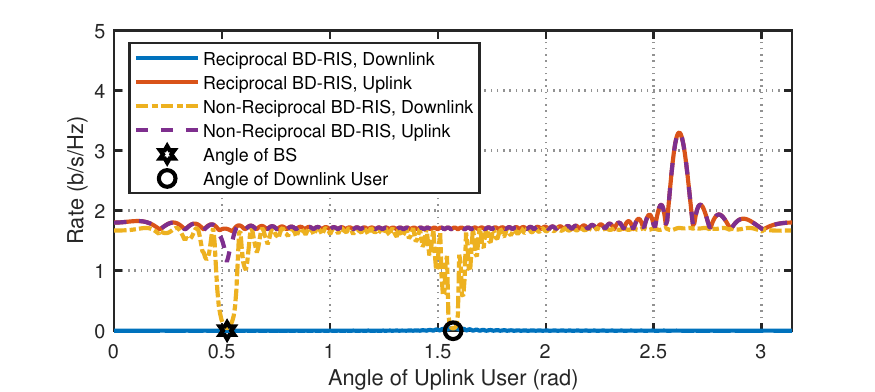}}
    \subfigure[][$\phi_{R_DI} = \frac{5\pi}{6}$]{\includegraphics[width = 0.48\textwidth]{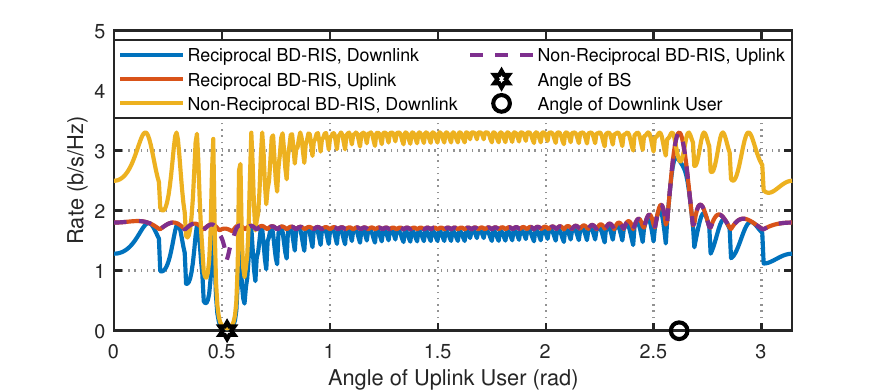}}
    \caption{The uplink and downlink rates versus the location of the uplink user using reciprocal and non-reciprocal BD-RISs with structural scattering ($\phi_{BI} = \frac{\pi}{6}$, $N_I=64$).}
    \label{fig:rate}
\end{figure}

\subsection{Uplink and Downlink Rates}

\subsubsection{No Structural Scattering}
Fig. \ref{fig:rate_noss} shows the uplink rate in (\ref{eq:rate_uplink_noss}) and downlink rate in (\ref{eq:rate_downlink_noss}) versus the location of the uplink user using reciprocal and non-reciprocal BD-RISs, from which we have the following observations. 
\textit{First}, when the uplink user is aligned with the base station, i.e., $\phi_{IT_U} = \phi_{BI}$, the non-reciprocal BD-RIS achieves nearly zero downlink rate due to the interference in (\ref{eq:rate_downlink_noss}), i.e., $|\mathbf{h}_{R_DI}^\mathsf{T}\mathbf{\Theta}\mathbf{h}_{IT_U}|^2$, which is as strong as the power of the signal of interest in (\ref{eq:rate_downlink_noss}), i.e., $|\mathbf{h}_{R_DI}^\mathsf{T}\mathbf{\Theta}\mathbf{h}_{BI}|^2$. 
\textit{Second}, when the uplink user is in other locations, the trend aligns with that in Fig. \ref{fig:channel_strength_noss}, which, again, verifies the discussion in Remark 3. 

\subsubsection{Structural Scattering}
Fig. \ref{fig:rate} shows the uplink rate in (\ref{eq:rate_uplink}) and downlink rate in (\ref{eq:rate_downlink}) versus the location of the uplink user using reciprocal and non-reciprocal BD-RISs, with structural scattering captured in the modeling. 
In addition to the conclusions obtained from Fig. \ref{fig:rate_noss}, we have the following observations.
\textit{First}, as shown in Fig. \ref{fig:rate}(a), when the uplink user is aligned with the downlink user with $\phi_{R_DI} = \phi_{IT_U} = \frac{\pi}{2}$, both non-reciprocal and reciprocal BD-RISs achieve nearly zero downlink rates. 
This is because the power of the structural scattering, i.e., $|-\mathbf{h}_{R_DI}^\mathsf{T}\mathbf{h}_{IT_U}|^2$, achieves its maximum with $\phi_{R_DI} + \phi_{IT_U} = \pi$, which results in strong interference in (\ref{eq:rate_downlink}).
However, in the case where the structural scattering at BD-RIS is ignored as shown in Fig. \ref{fig:rate_noss}, when the uplink user is aligned with the downlink user, both non-reciprocal and reciprocal BD-RIS achieve maximum rates due to the negligible interference. 
\textit{Second}, as shown in Fig. \ref{fig:rate}(a), the reciprocal BD-RIS fails to serve the downlink user when focusing only on the uplink user, while the non-reciprocal BD-RIS can simultaneously serve both downlink and uplink with proper deployments of devices. 
Meanwhile, in the case where the structural scattering at BD-RIS is ignored, the reciprocal BD-RIS can only serve two users when they are aligned, while the non-reciprocal BD-RIS could simultaneously serve two users as long as they are not aligned with the base station. 
This, again, highlights the benefit of using non-reciprocal BD-RIS in full-duplex systems and the importance of accurately capturing the structural scattering. 
\textit{Third}, as shown in Fig. \ref{fig:rate}(b), when the downlink user is symmetrically located with base station with $\phi_{R_DI} = \pi-\phi_{BI}$, the reciprocal BD-RIS can simultaneously serve two users since the benefit of structural scattering is captured in the downlink. 

\subsubsection{Design without and Evaluation with Structural Scattering} 
\begin{figure}[t]
    \centering
    \subfigure[][$\phi_{R_DI} = \frac{\pi}{2}$]{\includegraphics[width = 0.48\textwidth]{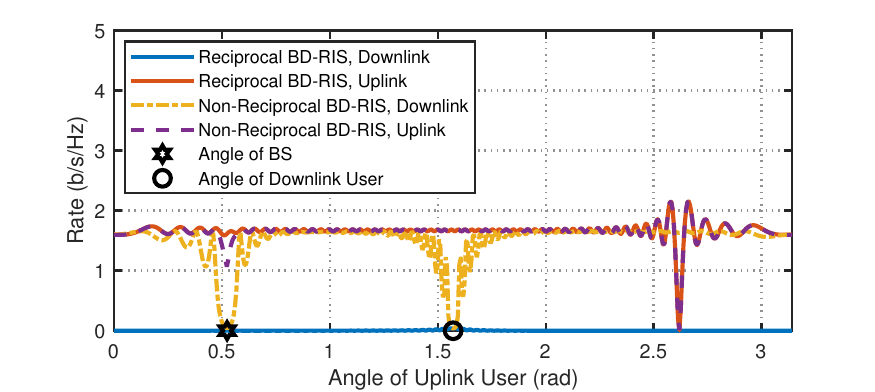}}
    \subfigure[][$\phi_{R_DI} = \frac{5\pi}{6}$]{\includegraphics[width = 0.48\textwidth]{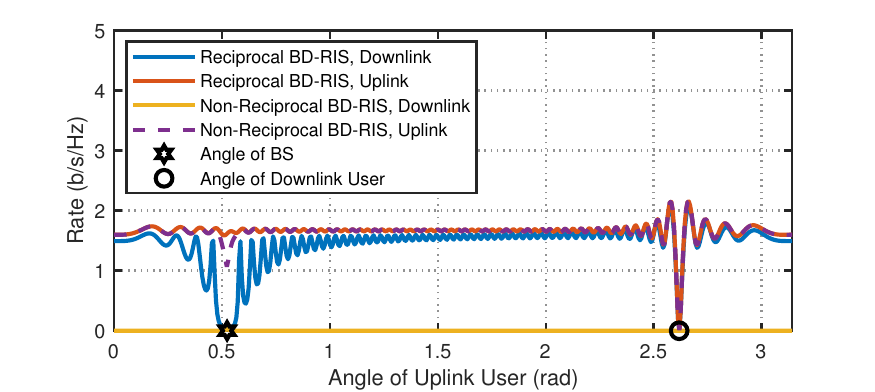}}
    \caption{The uplink and downlink rates versus the location of the uplink user using reciprocal and non-reciprocal BD-RISs without structural scattering for design and with for evaluation ($\phi_{BI} = \frac{\pi}{6}$, $N_I=64$).}
    \label{fig:rate_deswoss_evawss}
\end{figure}

Fig. \ref{fig:rate_deswoss_evawss} shows the uplink rate in (\ref{eq:rate_uplink}) and downlink rate in (\ref{eq:rate_downlink}) versus the location of the uplink user using reciprocal and non-reciprocal BD-RISs, with structural scattering captured for evaluation but ignored in the design. 
Comparing Fig. \ref{fig:rate_deswoss_evawss}(b) with Fig. \ref{fig:rate}(b), we observe the non-reciprocal BD-RIS fails to simultaneously serve two users while the reciprocal BD-RIS can. This can be explained following  the interpretation to Fig. \ref{fig:channel_strength_deswoss_evawss}. More importantly, this implies that the benefit of reciprocal BD-RIS in full-duplex systems can only be fully captured when the structural scattering is accurately included in the model and design. 

\section{Conclusion}
\label{sec:conclusion}

In this work, we studied the impact of BD-RIS reciprocity in full-duplex systems. To do so, we first derived the general RIS aided full-duplex system model using multiport network theory and simplified it based on physically consistent assumptions. Specifically, in the derived system model, the impact of structural scattering at BD-RIS was highlighted. With the simplified model, we identify a scenario where non-reciprocal BD-RIS could have significant gains over reciprocal BD-RIS. We showed the benefit of non-reciprocal BD-RIS by theoretically analyzing the conditions for reciprocal and non-reciprocal BD-RISs to simultaneously maximize the received power of the signal of interest at uplink and downlink. We finally provided simulation results to verify the theoretical derivations and to visualize the advantage of using non-reciprocal BD-RIS in full-duplex systems. Our results demonstrated that non-reciprocal BD-RIS has unique benefits in full-space systems since  its non-reciprocal property enables a full-duplex base station to optimally communicate with both downlink and uplink users at different locations. 

Inspired by the performance benefits of non-reciprocal BD-RIS in full-duplex systems, future research directions include but are not limited to 1) exploring the use of other BD-RIS architectures, e.g., simultaneously transmitting and reflecting (STAR) RIS \cite{xu2021star}, which could potentially enhance simultaneously the uplink and downlink transmissions in full-duplex systems; 2) exploring the benefit of non-reciprocal BD-RIS in simultaneously enhancing the performance for sensing and communications; 3) modeling and analyzing the power loss of non-reciprocal BD-RIS; 4) developing optimization methods based on the circuit design of non-reciprocal BD-RIS.

\begin{appendix}[Proof to Proposition 1]
    
The four properties in Proposition 1 are proven one by one as detailed below. 

\textit{Proof to Property 1:} 
Problem (\ref{eq:assym_opt}) can be re-written as 
\begin{equation}
    \begin{aligned}
        \mathbf{\Theta}^{\mathsf{NR}} &= \mathop{\arg\min}\limits_{\mathbf{\Theta}^\mathsf{H}\mathbf{\Theta}=\mathbf{I}_{N_I}}\|\mathbf{X}\|_\mathsf{F}^2 + \|\mathbf{Y}\|_\mathsf{F}^2 - 2\mathsf{Tr}(\Re\{\mathbf{X}\mathbf{Y}^\mathsf{H}\mathbf{\Theta}^\mathsf{H}\}),\\
        &= \mathop{\arg\max}\limits_{\mathbf{\Theta}^\mathsf{H}\mathbf{\Theta}=\mathbf{I}_{N_I}}\mathsf{Tr}(\Re\{\mathbf{X}\mathbf{Y}^\mathsf{H}\mathbf{\Theta}^\mathsf{H}\}),\\
        &= \mathop{\arg\max}\limits_{\mathbf{\Theta}^\mathsf{H}\mathbf{\Theta}=\mathbf{I}_{N_I}}\mathsf{Tr}(\Re\{\mathbf{U}\mathbf{\Sigma}\mathbf{V}^\mathsf{H}\mathbf{\Theta}^\mathsf{H}\}),\\
        &= \mathop{\arg\max}\limits_{\mathbf{\Theta}^\mathsf{H}\mathbf{\Theta}=\mathbf{I}_{N_I}}\mathsf{Tr}(\Re\{\mathbf{\Sigma}\mathbf{V}^\mathsf{H}\mathbf{\Theta}^\mathsf{H}\mathbf{U}\}),   
    \end{aligned}
\end{equation}
where the quantity $\mathbf{V}^\mathsf{H}\mathbf{\Theta}^\mathsf{H}\mathbf{U}$ is an unitary matrix and thus the maximum is achieved when $\mathbf{V}^\mathsf{H}\mathbf{\Theta}^\mathsf{H}\mathbf{U} = \mathbf{I}_{N_I\times N_I}$. This leads to the solution in (\ref{eq:opt_assym_theta}).

\textit{Proof to Property 2:} Plugging (\ref{eq:opt_assym_theta}) into (\ref{eq:assym_opt}) and applying $\|\mathbf{X}\|_\mathsf{F} = \|\mathbf{Y}\|_\mathsf{F} = \sqrt{2}$ give the global minimum in (\ref{eq:minimum}). 

\textit{Proof to Property 3:} The inequality can be found in \cite{hogben2006handbook}, while the equality of (b) holds when $\mathbf{X}$ and $\mathbf{Y}^\mathsf{H}$ have aligned singular vectors, i.e., there exist SVDs 
\begin{equation}
    \mathbf{X} = \mathbf{U}_{\mathbf{X}}\mathbf{\Sigma}_\mathbf{X}\bar{\mathbf{V}}^\mathsf{H}, ~\mathbf{Y}^\mathsf{H} = \bar{\mathbf{V}}\mathbf{\Sigma}_\mathbf{Y}\mathbf{V}_{\mathbf{Y}}^\mathsf{H},
\end{equation} 
with unitary matrices $\mathbf{U}_{\mathbf{X}}\in\mathbb{C}^{N_I\times N_I}$, $\mathbf{V}_{\mathbf{Y}}\in\mathbb{C}^{N_I\times N_I}$, and $\bar{\mathbf{V}}\in\mathbb{C}^{2\times 2}$.  
In this sense, the SVD of $\mathbf{X}\mathbf{Y}^\mathsf{H}$ can be written as 
\begin{equation}
    \mathbf{X}\mathbf{Y}^\mathsf{H} = \mathbf{U}_{\mathbf{X}}\mathbf{\Sigma}_\mathbf{X}\mathbf{\Sigma}_\mathbf{Y}\mathbf{V}_{\mathbf{Y}}^{\mathsf{H}},
\end{equation} 
implying that $\mathsf{Tr}(\mathbf{\Sigma}) = \mathsf{Tr}(\mathbf{\Sigma}_\mathbf{X}\mathbf{\Sigma}_\mathbf{Y})$. 

\textit{Proof to Property 4:}
By $\|\mathbf{X}\|_\mathsf{F} = \sqrt{2}$ and $\|\mathbf{Y}\|_\mathsf{F} = \sqrt{2}$, we have that 
\begin{equation}
    \begin{aligned}
        &\sigma_{\mathbf{X},1}^2 + \sigma_{\mathbf{X},2}^2 = \mathsf{Tr}(\mathbf{X}\mathbf{X}^\mathsf{H}) = 2,\\ 
        &\sigma_{\mathbf{Y},1}^2 + \sigma_{\mathbf{Y},2}^2 = \mathsf{Tr}(\mathbf{Y}\mathbf{Y}^\mathsf{H}) = 2. 
    \end{aligned}
\end{equation}
Using the Cauchy-Schwarz inequality, we have 
\begin{equation}
    \begin{aligned}
    &(\sigma_{\mathbf{X},1}\sigma_{\mathbf{Y},1} + \sigma_{\mathbf{X},2}\sigma_{\mathbf{Y},2})^2\\
    &\le (\sigma_{\mathbf{X},1}^2 + \sigma_{\mathbf{X},2}^2)(\sigma_{\mathbf{Y},1}^2 + \sigma_{\mathbf{Y},2}^2) = 4,
    \end{aligned}
\end{equation}
where the equality holds when $[\sigma_{\mathbf{X},1},\sigma_{\mathbf{X},2}]^\mathsf{T}$ and $[\sigma_{\mathbf{Y},1},\sigma_{\mathbf{Y},2}]^\mathsf{T}$ are linearly dependent, i.e., $[\sigma_{\mathbf{X},1},\sigma_{\mathbf{X},2}]^\mathsf{T} = \rho[\sigma_{\mathbf{Y},1},\sigma_{\mathbf{Y},2}]^\mathsf{T}$. 
Meanwhile, by $\sigma_{j,1}\ge\sigma_{j,2}$ and $\sigma_{j,1}^2 + \sigma_{j,2}^2 = 2$, we have 
$1\le\sigma_{j,1}\le\sqrt{2}$, $\forall j\in\{\mathbf{X},\mathbf{Y}\}$, which indicates that $\rho = 1$.
In addition, we have
\begin{equation}
    \begin{aligned}
    &(\sigma_{\mathbf{X},1}\sigma_{\mathbf{Y},1} + \sigma_{\mathbf{X},2}\sigma_{\mathbf{Y},2})^2\\ 
    &= 2(\sigma_{\mathbf{X},1}^2 - 1)(\sigma_{\mathbf{Y},1}^2 - 1) + 2\sigma_{\mathbf{X},1}\sigma_{\mathbf{Y},1}\sqrt{2 - \sigma_{\mathbf{X},1}^2}\\
    &~~~\times\sqrt{2-\sigma_{\mathbf{Y},1}^2} + 2 \ge 2,
    \end{aligned}
\end{equation} 
where the equality holds when the first two terms equal to zero simultaneously, i.e., $\sigma_{\mathbf{X},1} = 1$ and $\sigma_{\mathbf{Y},1}=\sqrt{2}$, or $\sigma_{\mathbf{Y},1} = 1$ and $\sigma_{\mathbf{X},1}=\sqrt{2}$.

The proofs are completed. $\hfill\square$
\end{appendix}

\bibliographystyle{IEEEtran}
\bibliography{refs}

\end{document}